\documentclass[12pt,preprint,preprintnumbers,nofootinbib]{revtex4}

\usepackage{graphicx}
\usepackage{epsfig}    
\usepackage{dcolumn} 
\usepackage{bm}      
\usepackage{amssymb} 
\usepackage{epsfig}    
\usepackage{here}

\newcommand{\nn}{\nonumber}

\newcommand{\al}{\alpha}
\newcommand{\be}{\beta}

\newcommand{\gsim}{\mbox{ \raisebox{-1.0ex}{$\stackrel{\textstyle >}
{\textstyle \sim}$ }}}
\newcommand{\lsim}{\mbox{ \raisebox{-1.0ex}{$\stackrel{\textstyle <}
{\textstyle \sim}$ }}}

\def\Journal#1#2#3#4{{#1} {\bf #2} (#4) #3}

\def\NPB{{\em Nucl. Phys.} B}
\def\PLB{{\em Phys. Lett.} B}
\def\PRL{\em Phys. Rev. Lett.}
 
\def\PRD{{\em Phys. Rev.} D}
\def\ZPC{{\em Z. Phys.} C} 
\def\EPC{{\em Euro. Phys. J.} C}
\def\PTP{\em Prog.~Theor.~Phys.}

\begin{document}
\topmargin -2cm

\begin{flushright}

\today
\end{flushright}

\begin{center}
{\large \bf 
Higgs coupling constants as a probe of new physics}

\vspace*{4mm}
{{\sc Shinya Kanemura}$^{a,\,}$\footnote{
                   E-mail: kanemu@het.phys.sci.osaka-u.ac.jp}, 
        {\sc Yasuhiro Okada}$^{b,c,\,}$\footnote{
                   E-mail: yasuhiro.okada@kek.jp},  
        {\sc Eibun Senaha}$^{b,c,\,}$\footnote{E-mail: senaha@post.kek.jp} \\
    and    {\sc C.-P. Yuan}$^{d,\,}$\footnote{E-mail: yuan@pa.msu.edu}}    

{
{\em $^a$Department of Physics, Osaka University, Toyonaka, Osaka
 560-0043, Japan\\
     $^b$Theory Group, KEK, Tsukuba, Ibaraki 305-0801, Japan\\
     $^c$Department of Particle and Nuclear Physics, 
         the Graduate University for Advanced Studies, Tsukuba, 
         Ibaraki 305-0801, Japan\\ 
     $^d$Department of Physics and Astronomy, Michigan State University, 
     East Lansing, Michigan 48824-1116, USA}}
\end{center}

\begin{abstract}
We study new physics effects on the couplings of weak gauge bosons 
with the lightest CP-even Higgs boson ($h$), $hZZ$, and 
the tri-linear coupling of the lightest Higgs boson, $hhh$, 
at the one loop order, as predicted by the two Higgs doublet model.  
Those renormalized coupling constants can deviate from the Standard 
Model (SM) predictions due to two distinct origins;  
the tree level mixing effect of Higgs bosons 
and the quantum effect of additional particles in loop diagrams.   
The latter can be enhanced in the renormalized $hhh$ coupling 
constant when the additional particles show the non-decoupling property. 
Therefore, even in the case where the $hZZ$ coupling is close to 
the SM value, deviation in the $hhh$ coupling from the SM value 
can become as large as plus $100$ percent, 
while that in the $hZZ$ coupling is at most minus $1$ percent level.
Such large quantum effect on the Higgs tri-linear coupling is 
distinguishable from the tree level mixing effect, and 
is expected to be detectable at a future linear collider. 
\end{abstract}

\maketitle

\section{Introduction}

In the standard picture of elementary particle physics, 
the electroweak gauge symmetry is spontaneously broken by 
introducing an iso-doublet scalar field, the Higgs field. 
Its neutral component receives the vacuum expectation value.  
Consequently, the gauge bosons and the matter fields obtain 
their masses through the couplings with the Higgs scalar field. 
 
Identification of the Higgs boson 
is one of the most important goals of high energy collider experiments. 
The fit by LEP Electroweak Working Group favors a relatively light Higgs boson 
with its mass below 251 GeV, assuming the Standard Model (SM)\cite{LEPEWWG}.
The search for the Higgs bosons is being carried at Fermilab
Tevatron and will be continued at CERN Large Hadron Collider (LHC).
There the SM Higgs boson is expected 
to be discovered as long as its mass is less than 1 TeV. 
In order to verify the mechanism of mass generation, the Higgs boson 
couplings with gauge bosons as well as fermions have to be determined 
with sufficient accuracy. Moreover, precise determination of 
the self-coupling constant of the Higgs boson is essential to 
determine the structure of the Higgs potential.
An electron-positron ($e^-e^+$) linear collider (LC), such as 
GLC\cite{LC-glc}, 
TESLA\cite{LC-tesla} or NLC\cite{LC-nlc} and its photon-photon 
($\gamma$-$\gamma$)
collider option, can provide an opportunity for the precise measurement 
of the Higgs boson couplings. 
At LC's, the Higgs boson ($h$) is produced mainly via the 
Higgsstrahlung process $e^+e^- \to Z h$
for relatively low energies and also via the fusion process   
$e^+e^- \to W^{+\ast}W^{-\ast} \nu \bar\nu \to h \nu \bar\nu$ 
for higher energies\cite{eehx}. 
In both production mechanisms, the Higgs boson is produced 
through the coupling with weak gauge bosons.  
The cross sections are expected to be measured at a percent level or 
better unless the Higgs boson is relatively heavy. 
The Higgs boson couplings with heavy quarks (except the top quark) 
and the tau lepton 
can be tested by measuring the decay branching 
ratios of the Higgs boson. 
Furthermore, the tri-linear coupling of the Higgs boson 
$hhh$\cite{battaglia,hhh_yamashita,hhh_ACFA,hhh_LHC,
hhh_ext,dec-region,hhh_hollik,lcws02kkosy,kkosy,barger} 
and the top-Yukawa coupling $h t \bar t$
can be determined by measuring the cross section 
of double Higgs production processes\cite{eehhx,eehhx2,eehhx3} 
$e^+e^- \to Z h h$ as well as  
$e^+e^- \to W^{+\ast}W^{-\ast} \nu \bar\nu \to h h \nu \bar\nu$ 
and the top-associated Higgs production process\cite{htt}
$e^+e^- \to  h t \bar t$, respectively. 
The $\gamma\gamma$ option of the LC can also be useful for 
the Higgs self-coupling measurement\cite{jikia}.

Studying the Higgs sector is not only useful for the 
confirmation of the breaking mechanism of the electroweak gauge
symmetry,  but also provides a sensitive window for new physics 
beyond the SM. 
In fact, in many models of new physics an extended Higgs sector 
appears as the low energy effective theory, which has discriminative 
phenomenological properties.
One popular example is known to be the minimal supersymmetric standard 
model (MSSM)\cite{HHG}, 
in which the Higgs sector is a two Higgs doublet model (THDM). 
Some models of the dynamical breaking of the electroweak symmetry
also induce more than one Higgs doublets in their low energy 
effective theories\cite{TC}. 
There are other motivations to introduce extra Higgs fields, 
such as electroweak baryogenesis\cite{baryogenesis}, 
top-bottom mass hierarchy\cite{top-bottom}, and
neutrino mass problem\cite{zee}.

A common feature of extended Higgs sectors is the existence 
of additional scalar bosons, such as charged Higgs bosons and 
CP-odd Higgs boson(s). 
After the discovery of the lightest Higgs boson,  
direct search of these extra scalar particles would
become important to distinguish new physics models from 
the SM. Even if the extra Higgs bosons are not found,  
we can still obtain insight by looking for indirect effects of 
the extra Higgs boson from the precise determination of  
the lightest Higgs boson properties\cite{LC-indirect}. 
For example, the mass, width, production cross sections 
and decay branching ratios of the lightest Higgs boson
should be thoroughly measured 
to test whether or not these data are consistent with the SM.  
The existence of extra Higgs bosons can affect the observables 
associated with the lightest Higgs boson 
through both the tree level mixing effect and the quantum loop effect. 
In this way, we might find clues to new physics  
before finding the extra Higgs bosons from the direct search experiments. 

In this paper, we evaluate the Higgs coupling 
with the gauge boson $hZZ$ and the Higgs self-coupling $hhh$ 
at one loop level in the THDM, in order to study 
the impact of the extra Higgs bosons on the 
coupling associated with the lightest Higgs boson ($h$).  
In Refs.~\cite{lcws02kkosy,kkosy}, the one loop contributions of additional 
Higgs bosons and heavy quarks to the $hhh$ coupling are discussed 
in the limit where only $h$ is responsible for the 
electroweak symmetry breaking (in the SM-like limit). 
The calculation has been done both in the 
effective potential method and in the diagrammatic method, but 
details of the calculation were not shown.
In the present paper, we will show the details of our calculation, 
in which the on-shell renormalization scheme\cite{on-shell} is 
adopted. In addition, new particle effects on the form factors of 
the $hZZ$ coupling are also evaluated.
Furthermore, we also extend our discussion in
Refs.~\cite{lcws02kkosy,kkosy} for the case of the SM-like limit 
to more generic cases.

In the THDM, masses of the heavy Higgs bosons 
can come from two kinds of contributions: 
the vacuum expectation value $v$ ($\simeq 246$ GeV) 
and the gauge invariant mass term.
When the heavy Higgs boson mass is predominantly generated by $v$, 
contributions in powers of the mass of the loop particles 
can appear in the one loop effect couplings of $hZZ$ and $hhh$. 
They are quadratic for the $h ZZ$ coupling and quartic 
for the $hhh$ coupling\cite{lcws02kkosy,kkosy}. 
In this case, relatively large quantum correction is expected in 
the $hhh$ vertex, especially when the particle in the loop is heavy.
Although similar non-decoupling loop effects can also appear 
in the THDM\cite{nondec_THDM} 
in the processes of 
$h \to \gamma\gamma$\cite{hgammagamma,hbbbar}, 
$h \to b \bar b$\cite{hbbbar}, 
$e^+e^- \to W^+W^-$\cite{eeww} and those with 
the coupling $W^\pm H^\mp V$ ($V=Z,\gamma$)\cite{whz}, 
the quartic power contribution of the mass is a unique feature 
of the $hhh$ coupling. 
These observables can receive large quantum corrections 
due to the non-decoupling effects.
On the contrary, when the heavy Higgs bosons obtain their masses 
mainly from the other source, 
such power-like contribution disappears and the one loop effects vanish
in the large mass limit.
The Higgs sector of the MSSM belongs to this case.

At the tree level, both the $hZZ$ and $hhh$ coupling constants 
of the THDM can largely deviate from the SM values due to the 
Higgs mixing effect. 
The $hZZ$ coupling is given by the multiplication 
of the factor $\sin(\be-\al)$ to the SM coupling constant, where 
$\tan\beta$ 
is the ratio of the vacuum expectation values and $\alpha$ is the mixing 
angle between CP-even Higgs bosons. 
In the limit of $\sin(\be-\al) = 1$, where the $hZZ$ coupling recovers  
the SM value, the $hhh$ coupling also approaches to the SM prediction 
for a given mass of the Higgs boson $h$. 
We study how this correlation can be changed by the one loop corrections.

We evaluate the one loop effects due to additional Higgs bosons 
as well as the top quark under the constraint from 
the perturbative unitarity\cite{LQT,unitarity1,unitarity2}
and the vacuum stability\cite{vacuum-stability}.  
The constraint from the available precision data such as the $\rho$
parameter constraint is also taken into account\cite{rhoTHDM,ST_2HDM}.  

The one loop effect on the $hZZ$ coupling can be 
as large as (minus) one percent of the SM coupling  
in the wide range of parameter space.   
This shows that a larger negative deviation 
can only be realized due to the Higgs mixing effect, 
i.e., the effect of the factor $\sin(\be-\al)$. 
If the observed $hZZ$ coupling agrees with the SM prediction 
within the 1\% accuracy, 
we may not be able to distinguish the quantum effect from 
the tree level mixing effect.

The deviation in the $hhh$ coupling can be 
as large as (plus) $100$ percent for the mass of $h$ to be around 120 GeV 
due to the non-decoupling quantum effect of the heavy extra Higgs bosons. 
This happens even in the SM-like limit, $\sin(\be-\al) \to 1$. 
Such magnitude of the deviation is larger than the experimental 
accuracy that is expected to be $10$-$20$ \% 
at LC's\cite{battaglia,hhh_yamashita,hhh_ACFA}, and can be 
experimentally detected.
Therefore, the combination of precise measurements 
of the $h ZZ$ and $hhh$ couplings can be useful to explore 
the structure of the Higgs sector.

In Sec.~\ref{sec:2}, the form factors of $h ZZ$ coupling and 
$hhh$ coupling are defined, and the SM contribution 
to them is briefly discussed. 
In Sec.~\ref{sec:3}, the general feature of the THDM is summarized, and 
the renormalization scheme of the THDM is defined. 
The one loop contributions to the form factors of 
the $hZZ$ and $hhh$ couplings are calculated in Sec.~\ref{sec:4}.
The analytic properties of the loop corrections are discussed in
Sec.~\ref{sec:5},
and the numerical evaluation is shown in Sec.~\ref{sec:6}. 
Sec.~\ref{sec:7} contains our conclusions. 
For completeness, we also present the details of our calculation 
in the Appendices.

\section{The $hZZ$ and $hhh$ couplings in the SM}\label{sec:2}

Before showing the calculation of the form factors in the THDM, 
it is instructive to discuss the top-quark loop effect on the $hZZ$ 
and $hhh$ couplings in the SM. 
One can find a simple example of the non-decoupling effect 
in the top-quark loop contribution. It is also useful as technical 
introduction to the calculation in the THDM. 
In Appendix~\ref{app:SM}, we show details of 
the top-quark one loop contribution 
to the $hZZ$ and $hhh$ couplings in the SM.

The most general form factors of the $hZZ$ coupling can be written as
\begin{eqnarray}  
 M^{\mu\nu}_{hZZ}= M_1^{hZZ} g^{\mu\nu} 
                 + M_2^{hZZ} \frac{p_1^\nu p_2^\mu}{m_Z^2}
                 + M_3^{hZZ} \,i \epsilon^{\mu\nu\rho\sigma} 
                           \frac{{p_1}_\rho^{} {p_2}_\sigma^{}}{m_Z^2}, 
\label{eq:hzz-form}
\end{eqnarray}
where $m_Z^{}$ is the mass of the Z boson, 
$p_1$ and $p_2$ are the momenta of incoming $Z$ bosons, and we define 
$g^{\mu\nu}={\rm diag}(1,-1,-1,-1)$ and 
$\epsilon^{0123}=-1$. 
The tri-linear $hhh$ coupling of the Higgs boson 
is parameterized by 
\begin{eqnarray}
  {\cal L}_{\rm self} 
  =    +   \frac{1}{3 !}\Gamma_{hhh}   h^3. 
\end{eqnarray}
In the SM, the lowest order contributions to the form factors 
$M_{1{\rm -}3}^{hZZ}$ and $\Gamma_{hhh}$ are given by 
\begin{eqnarray}
&& M_1^{hZZ(tree)} 
  = \frac{2 m_Z^2}{v}, \,\,\,\, M_2^{hZZ(tree)} = M_3^{hZZ(tree)} =0, \\
&& \Gamma_{hhh}^{tree} = -\frac{3 m_h^2}{v}, 
\end{eqnarray}
where $m_h^{}$ is the mass of the Higgs boson.

Let us consider the loop contribution of the top quark to these 
form factors. Details of calculation are presented in 
Appendix~\ref{app:SM}.
From the naive power counting, it is understood that 
$M_1^{hZZ}$ receives the highest power contribution of the 
top quark mass among the form factors of $hZZ$ vertex 
$M_i^{hZZ}$ ($i=1-3$).   
The leading one loop contribution of the top quark to the 
form factor $M_1^{hZZ}$ is calculated as 
\begin{eqnarray}
  M_{1}^{hZZ}(p_1^2,p_2^2,p_3^2) &=& \frac{2 m_Z^2}{v} 
  \left[ 1 - \frac{1}{16\pi^2}  \frac{5}{2} \frac{m_t^2}{v^2} 
  \left\{ 1 
  + {\cal O}\left(\frac{m_h^2}{m_t^2},\frac{p_i^2}{m_t^2}\right) 
 \right\}\right], 
 \label{sm_hzz_top}
\end{eqnarray}
where $m_t$ is the mass of the top quark, $p_i$ ($i=1-3$) represent 
the momenta of the external lines.
The leading top quark contribution to $M_{1}^{hWW}$ is 
the same as that to $M_{1}^{hZZ}$, because of the isospin symmetry. 
Both have the quadratic power 
contribution of the top quark mass.
On the other hand, the leading contribution of the top quark 
to the self-coupling constant is calculated\cite{kkosy} as
\begin{eqnarray}
  \Gamma_{hhh}^{}(p_1^2,p_2^2,p_3^2) &=& - \frac{3 m_h^2}{v} 
  \left[ 1 
- \frac{1}{16 \pi^2}  \frac{16 m_t^4}{v^2 m_h^2}
  \left\{ 1 
  + {\cal O}\left(\frac{m_h^2}{m_t^2},\frac{p_i^2}{m_t^2}\right) 
 \right\} \right]. 
 \label{sm_hhh_top}
\end{eqnarray}
The top-quark contribution is quartic in mass, so that we expect 
larger corrections to the $hhh$ coupling than the 
correction to $hZZ$ vertices 
by the enhancement factor of $(32/5) m_t^2/m_h^2$ especially 
when $m_h < m_t$. 
Eq.~(\ref{sm_hhh_top}) shows that the leading contribution of 
the top quark loop deviates the $hhh$ form factor from the 
tree level value by about $-12$\% for $m_t=178$ GeV and $m_h=120$ GeV. 
The quartic dependence of the top quark mass is also reproduced 
easily in the effective potential method as shown in 
Appendix~\ref{app:SMeff}. 

At the future collider experiment, the $hhh$ coupling will be measured 
via the double Higgs production processes, where at least 
one of the three legs of the $hhh$ vertex is off-shell.  
Thus the momentum dependence in the $hhh$ form factor 
is important.  
In Fig.~1, the top quark loop contribution to the effective 
$hhh$ coupling $\Gamma_{hhh}(q^2)$ is shown 
as a function of the invariant mass ($\sqrt{q^2}$) of the virtual 
$h$ boson for $m_h^{}=100$, $120$ and $160$ GeV. 
$\Gamma_{hhh}(q^2)(\equiv
\Gamma_{hhh}(m_h^2,m_h^2,q^2))$  
is evaluated from Eq.~(\ref{sm_hhh}) in 
Appendix~\ref{app:SM1}.
In the small $\sqrt{q^2}$ limit ($\sqrt{q^2} \to 0$), 
the correction due to the top quark loop is negative 
and approaches to the similar value estimated 
from Eq.~(\ref{sm_hhh_top}).  
However, such a value of $\sqrt{q^2}$ is lower than the the 
threshold $2 m_h^{}$ of the subprocess $h^\ast \to hh$, and 
kinematically not allowed.
We find that the top-quark loop effect strongly 
depends on $\sqrt{q^2}$, because the threshold enhancement 
at $\sqrt{q^2}=2 m_t^{}$ contributes an opposite sign 
to the quartic mass term contribution.  
The correction changes the sign 
when $\sqrt{q^2}$ is somewhere between $2 m_h$ and $2 m_t$.  
The enhancement due to the top-pair threshold is maximum 
at the point just after the threshold of the top pair production.

\begin{figure}[t]
\vspace*{-12mm}
\begin{center}
\hspace*{-3mm}
\includegraphics[width=10cm,height=8cm]{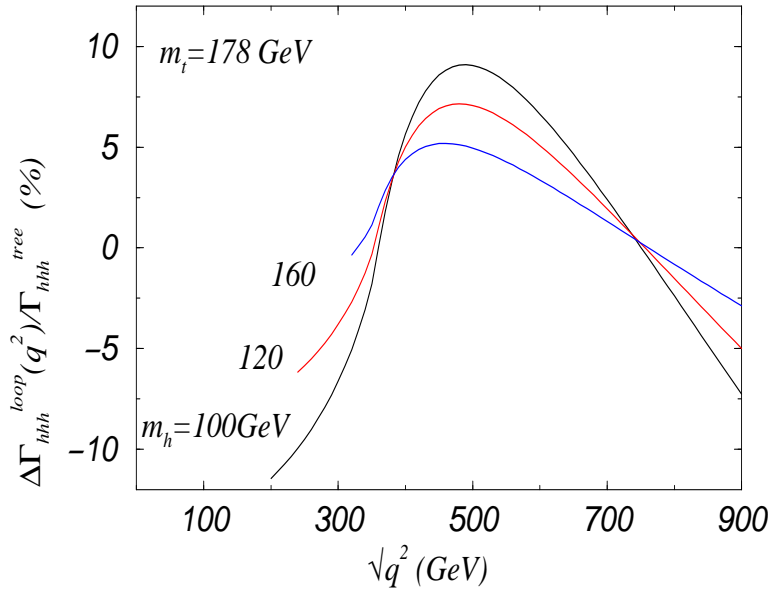}
\end{center}
\vspace*{-8mm}
\caption{     The one-loop contribution of the top quark 
              to the effective $hhh$ coupling as a function of 
              $\sqrt{q^2}$, where $q^\mu$ is the momentum 
              of the off-shell $h$ boson in $h^\ast \to hh$.  
              $\Delta\Gamma_{hhh}^{loop}(q^2)$ is defined by 
              $\Gamma_{hhh}(q^2)-\Gamma_{hhh}^{tree}$ 
              in the SM. } \label{fig:newSM2.eps}
\end{figure}

\section{The two Higgs doublet model}\label{sec:3}

In this section, we give a brief review of the THDM to make 
our notation clear and to prepare some tree level formulas,  
which will be used for the one loop calculation in the next section. 
We consider the model with a (softly-broken) discrete symmetry 
under the transformation $\Phi_1 \to \Phi_1$ and $\Phi_2 \to - \Phi_2$,
where $\Phi_i$ are the Higgs iso-doublets with hypercharge $\frac{1}{2}$.  
This discrete symmetry ensures the natural suppression of 
flavor changing neutral current at tree level.   
Two types of Yukawa interaction are then possible; 
i.e., so-called Model I and Model II\cite{HHG}. 
In Model I, only $\Phi_2$ is responsible for generating masses for 
all quarks and charged leptons, whereas in Model II $, \Phi_1 $ 
generates masses of down-type quarks and charged leptons and 
$ \Phi_2$ gives masses of up-type quarks.  
In our analysis, Model II Yukawa interaction is assumed throughout 
this paper. 
Later, we will comment on the case that Model I is considered. 

The Higgs potential is given by\cite{HHG} 
\begin{eqnarray}
  {V}_{\rm THDM}  &=&  m_1^2 \left| \Phi_1 \right|^2 
                          + m_2^2 \left| \Phi_2 \right|^2 - 
                              m_3^2 \left( \Phi_1^{\dagger} \Phi_2 
                                + \Phi_2^{\dagger} \Phi_1 \right) 
                                \nn  
                       + \frac{\lambda_1}{2} 
                               \left| \Phi_1 \right|^4 
                             + \frac{\lambda_2}{2} 
                               \left| \Phi_2 \right|^4 \nn \\
& &                          + \lambda_3 \left| \Phi_1 \right|^2 
                                \left| \Phi_2 \right|^2 
                      + \lambda_4 
                               \left| \Phi_1^{\dagger} \Phi_2 \right|^2
                             + \frac{\lambda_5}{2} 
                             \left\{ 
                               \left( \Phi_1^{\dagger} \Phi_2 \right)^2
                            +  \left( \Phi_2^{\dagger} \Phi_1 \right)^2
                             \right\},  \label{pot}
\end{eqnarray}
where $m_1^2$, $m_2^2$ and $\lambda_1$ to $\lambda_4$ are real, 
while $m_3^2$ and $\lambda_5$ are generally complex. 
We here assume that there is no CP violation in the Higgs sector,  
so as to neglect the phases of $m_3^2$ and $\lambda_5$. 
A nonzero value of $m_3^2$ indicates that the discrete symmetry 
is broken softly. Under the assumption, 
there are eight real parameters in the potential 
(\ref{pot}). The Higgs sector of the MSSM is a special case of 
Eq.~(\ref{pot}) with Model II Yukawa interaction at tree level.

The Higgs doublets are parameterized as  
\begin{eqnarray}
  \Phi_i = \left[\begin{array}{c}
            w^+_i \\
            \frac{1}{\sqrt{2}}(v_i + h_i + i z_i)
         \end{array}\right],  \;\;\;(i=1,2), 
\end{eqnarray}
where $v_{i}$ ($i=1,2$)  are vacuum expectation values that 
satisfy $\sqrt{v_1^2+v_2^2} = v \simeq 246$ GeV.
We here assume the case with $v_1 v_2 \neq 0$. 
From the vacuum condition (the stationary condition), we obtain   
\begin{eqnarray}
   0 &=& m_3^2 v_2 - m_1^2 v_1 
           - \frac{1}{2} \lambda_1 v_1^3 
           - \frac{1}{2} 
 (\lambda_3+\lambda_4+\lambda_5) v_1 v_2^2, \\
   0 &=& m_3^2 v_1 - m_2^2 v_2 
           - \frac{1}{2} \lambda_2 v_2^3 
           - \frac{1}{2} 
 (\lambda_3+\lambda_4+\lambda_5) v_1^2 v_2, 
\end{eqnarray}
and the mass parameters $m_1^2$ and $m_2^2$
can be eliminated with their degrees of freedom being replaced 
by those of $v_1$ and $v_2$. 
The mass matrices of the Higgs bosons are diagonalized 
by introducing the mixing angles $\beta$ and $\alpha$.
First, we rotate the fields by $\beta$ as 
\begin{eqnarray}
    \left( \begin{array}{c}
       h_1 \\
       h_2
    \end{array}\right) 
    &=& R(\beta)
    \left( \begin{array}{c}
       \phi_1 \\
       \phi_2
    \end{array}\right), 
    \left( \begin{array}{c}
       z_1 \\
       z_2
    \end{array}\right) 
    =
     R(\beta)
    \left( \begin{array}{c}
       z \\
       A
    \end{array}\right), \,\,
    \left( \begin{array}{c}
       w_1^+ \\
       w_2^+
    \end{array}\right) 
    =
    R(\beta)
    \left( \begin{array}{c}
       w^+ \\
       H^+
    \end{array}\right), \\
{\rm with} &&\,\,\,\, R(\theta) =
    \left( \begin{array}{cc}
       \cos\theta & -\sin\theta \\
       \sin\theta & \cos\theta
    \end{array}\right).  
\end{eqnarray}
By setting $\tan\beta=v_2/v_1$, the CP-odd and charged states are diagonalized.  
The Nambu-Goldstone bosons $z$ and $w^\pm$ are massless if 
the gauge interaction is switched off, and their degrees of 
freedom are eaten by the longitudinal components 
of $Z$ and $W^\pm$ bosons when the gauge interaction is turned on.
The masses of the physical states $A$ (CP-odd) and $H^\pm$ (charged) 
are expressed by 
\begin{eqnarray}
 m_{H^\pm}^2 &=& M^2 - \frac{1}{2} (\lambda_4+\lambda_5) v^2,\\
 m_{A}^2 &=& M^2 - \lambda_5 v^2, 
\end{eqnarray}
where $v$ is defined by $v=\sqrt{v_1^2+v_2^2}$, and 
$M$ is defined from the remaining degree of freedom 
of the mass parameter $m_3^2$ by $M^2 = m_3^2/\sin\beta\cos\beta$. 
The CP even states are not yet diagonalized, and the mass matrix
for $\phi_{1,2}$ is given by $M_{ij}^2$, where  
\begin{eqnarray}
 M_{11}^2 &=& (\lambda_1 \cos^4\beta + \lambda_2 \sin^4\beta 
                          + 2 \lambda \cos^2\beta\sin^2\beta) v^2 , \\
 M_{12}^2 &=& (- \lambda_1 \cos^2\beta + \lambda_2 \sin^2\beta 
                + \lambda \cos 2\beta) \cos\beta\sin\beta v^2, \\
 M_{22}^2 &=&  M^2 + \frac{1}{8} 
             (\lambda_1 + \lambda_2 - 2 \lambda) (1 - \cos 4\beta) v^2,   
\end{eqnarray}
with $\lambda=\lambda_3+\lambda_4+\lambda_5$.
The diagonalized CP-even states ($H$, $h$) are obtained 
from ($\phi_1$, $\phi_2$) 
by the rotation with the angle $(\alpha-\beta)$ as
\begin{eqnarray}
    \left( \begin{array}{c}
       \phi_1 \\
       \phi_2
    \end{array}\right) 
    &=& R(\alpha-\beta)
    \left( \begin{array}{c}
       H \\
       h
    \end{array}\right).  
\end{eqnarray}
The mixing angle $(\alpha-\beta)$ and the mass eigenstates 
are determined as 
\begin{eqnarray}
 \tan 2(\alpha-\beta) = 
  \frac{ 2 M_{12}^2} {M_{11}^2 - M_{22}^2}\,,
\end{eqnarray}
and
\begin{eqnarray}
 m_H^2 &=& \cos^2(\alpha-\beta) M_{11}^2 + \sin 2(\alpha-\beta) M_{12}^2 
           + \sin^2(\alpha-\beta) M_{22}^2, \\ 
 m_h^2 &=& \sin^2(\alpha-\beta) M_{11}^2 - \sin 2(\alpha-\beta) M_{12}^2 
           + \cos^2(\alpha-\beta) M_{22}^2, 
\end{eqnarray}
respectively.
The two physical CP-even fields $h$ and $H$ are defined so as to 
satisfy $m_h^{} \lsim m_H^{}$.
Among $M_{ij}^2$, only $M_{22}^2$ includes the dimensionful parameter 
$M$. In the limit of $M^2 \to \infty$, we have $\tan 2(\alpha-\beta) \to 0$. 
The angle $\alpha$ is chosen such that 
$m_h^2 \to M_{11}^2$, $m_H^2 \to M^2$ and 
$\sin(\alpha-\beta) \to -1$ are satisfied in the limit $M^2 \to + \infty$. 
\footnote{
Equivalently, we may rotate the CP-even fields from 
$(h_1, h_2)$ to $(H, h)$ by the angle $\alpha$ directly. 
Then we obtain 
\begin{eqnarray}
 \tan 2\alpha = \frac{\{ M^2 -(\lambda_3+\lambda_4+\lambda_5) v^2 \} 
                                                          \sin 2\beta }
                     { (M^2 - \lambda_1 v^2) \cos^2\beta 
                       - (M^2 - \lambda_2 v^2) \sin^2\beta } \,,
\end{eqnarray}
and 
\begin{eqnarray}
 m_H^2 &=&  M^2 \sin^2(\alpha-\beta) 
         + ( \lambda_1 \cos^2\alpha\cos^2\beta 
            +\lambda_2 \sin^2\alpha\sin^2\beta
            +\frac{1}{2} \lambda \sin 2\alpha \sin 2\beta ) v^2, \\ 
 m_h^2 &=&  M^2 \cos^2(\alpha-\beta) 
         + ( \lambda_1 \sin^2\alpha\cos^2\beta 
            +\lambda_2 \cos^2\alpha\sin^2\beta
            -\frac{1}{2} \lambda \sin 2\alpha \sin 2\beta ) v^2.
\end{eqnarray} 
One can easily check that the above two expressions for 
$m_h^{}$ and $m_H^{}$ are equivalent.
}

We note that the masses of the heavier Higgs bosons ($H$, $H^\pm$ and $A$) 
take the form as 
\begin{eqnarray}
  m_\Phi^2 = M^2 + \lambda_i v^2  \left( + {\cal O}(v^4/M^2)\right), 
\end{eqnarray}
where $\Phi$ represents $H$, $H^+$ or $A$ and 
$\lambda_i$ is a linear combination of $\lambda_1$-$\lambda_5$.
When $M^2 \gg \lambda_i v^2$, the mass $m_{\Phi}^2$ is determined 
by the soft-breaking scale of the discrete symmetry $M^2$, and 
is independent of $\lambda_i$.  
In this case, the effective theory below $M$ is described by 
one Higgs doublet, and all the tree level couplings 
related to the lightest Higgs boson $h$ approach 
to the SM value. 
Furthermore, the loop effects of $\Phi$ vanish 
in the large mass limit ($m_{\Phi}^{} \to \infty$) 
because of the decoupling theorem\cite{appelquist}.
The MSSM Higgs sector corresponds to this case, 
because $\lambda_i$ is fixed to be ${\cal O}(g^2)$ so that
large mass of $\Phi$ is possible only by large values of $M$. 
On the contrary, when $M^2$ is limited to be at the weak scale 
($M^2 \lsim \lambda_i v^2$) 
a large value of $m_\Phi^{}$ is realized by taking 
$\lambda_i$ to be large; i.e., the strong coupling regime. 
In this case, the squared mass of $\Phi$ is effectively proportional 
to $\lambda_i$, so that the decoupling theorem does not apply. 
Then, we expect a power-like contribution of $m_{\Phi}^{}$ 
in the radiative correction. 
We call such an effect as the non-decoupling effect of $\Phi$\cite{
nondec_THDM,hbbbar,hgammagamma,eeww,whz}. 
Similar non-decoupling effect appears in considering the top quark
loop contributions in the SM.
Although we expect large loop effects in this case, 
theoretical and experimental constraints must be considered. 
For instance, too large $\lambda_i$ leads to the 
breakdown of validity of perturbation 
calculation\cite{LQT,unitarity1,unitarity2}. 
Furthermore, the low energy precision data also impose 
important constraints on the model 
parameters\cite{peskin-takeuchi}. 
Later, in our evaluation of the one loop from factors, we shall take into 
account these constraints.

The parameters of the Higgs potential are $m_1^2$-$m_3^2$  
and $\lambda_1$-$\lambda_5$. 
They can be rewritten by eight ``physical'' parameters; 
i.e., four Higgs mass parameters $m_h, m_H, m_A, m_{H^\pm}$,  
two mixing angles $\al$, $\be$, the vacuum expectation value $v$, 
and the soft-breaking scale of the discrete symmetry $M$.   
The quartic coupling constants can be expressed in terms of these 
physical parameters as 
\begin{eqnarray}
  \lambda_1&=&\frac{1}{v^2 \cos^2\be} \left( - \sin^2\be M^2 
      + \sin^2\al m_h^2 + \cos^2\al m_H^2  \right),\label{mass_THDM1}\\
  \lambda_2&=&\frac{1}{v^2 \sin^2\be} \left( - \cos^2\be M^2
      + \cos^2\al m_h^2 + \sin^2\al m_H^2  \right),\label{mass_THDM2}\\
  \lambda_3&=&-\frac{M^2}{v^2}+2 \frac{m_{H^\pm}^2}{v^2}
              + \frac{1}{v^2} 
              \frac{\sin 2\al}{\sin 2\be}(m_H^2-m_h^2),\label{mass_THDM3}\\ 
  \lambda_4&=&\frac{1}{v^2} \left( 
      M^2 + m_A^2 - 2 m_{H^\pm}^2 \right),\label{mass_THDM4}\\
  \lambda_5&=&\frac{1}{v^2} \left( 
      M^2 - m_A^2 \right).\label{mass_THDM5}
\end{eqnarray}

\section{one loop correction to $hZZ$ and $hhh$ in the THDM}\label{sec:4}

We here discuss our scheme for calculating the one loop corrections to 
the form factors of $hZZ$ and $hhh$ in the THDM.  
As we are interested in the Higgs non-decoupling effects, 
we neglect the loop contributions of the gauge bosons in 
the calculation. This procedure is justified by adopting 
Landau gauge, where the effect of the gauge bosons and 
that of the Higgs bosons can be treated separately. 
The renormalization is performed in the on-shell 
scheme for physical mass parameters and mixing angles. 

First, we renormalize the three SM input parameters 
$m_W^{}$, $m_Z^{}$ and $G_F^{}$ ($=\frac{1}{\sqrt{2} v^2}$).
The counter terms of gauge boson masses 
($\delta m_W^2$, $\delta m_Z^2$) and the wave function 
renormalization factors ($\delta Z_W^{}$, $\delta Z_Z^{}$)
are obtained by calculating the transverse part $\Pi_T^{VV}(p^2)$ 
of the two-point function:
\begin{eqnarray}
 \Pi_{\mu\nu}^{VV}(p^2) = 
 \left(- g_{\mu\nu} + \frac{p_\mu p_\nu}{p^2}  \right) \Pi_T^{VV}(p^2) 
  + \frac{p_\mu p_\nu}{p^2}  \Pi_L^{VV}(p^2), 
\end{eqnarray}
where $VV=WW$ or $ZZ$.
In the on-shell renormalization scheme, we obtain 
\begin{eqnarray}
  \delta m_V^2 &=& {\rm Re} \Pi^{VV({\rm 1PI})}_T(m_V^2), \\ 
  \delta Z_V   &=& - \left.\frac{\partial}{\partial p^2 } 
                   {\rm Re} \Pi^{VV({\rm 1PI})}_T(p^2)
\right|_{p^2=m_V^2}. 
\end{eqnarray}
The renormalization for the vacuum expectation value $\delta v$ 
($v \to v + \delta v$) is defined by
\begin{eqnarray}
\frac{\delta v}{v} = \frac{1}{2}\,  \frac{1}{m_W^2} \Pi^{WW}_T(0) + 
 ({\rm vertex}\,\, {\rm and}\,\, {\rm box}\,\, {\rm corrections}).
\end{eqnarray}
When neglecting the vertex and box contributions, 
which are ${\cal O}(\alpha_{EM}^{})$, 
$\delta v/v$ can be expressed solely by the oblique correction 
$\Pi^{WW}_T(0)$.\footnote{
It is straight forward to see the difference 
from the other renormalization schemes in which 
the SM inputs are taken as $(\alpha,m_Z^{},G_F^{})$ or  
$(m_W^{},m_Z^{},\alpha_{EM}^{})$.
The difference is of order $\alpha$ which is neglected  
in the present calculation. 
} 

Next,
let us define the renormalization scheme for the Higgs sector.
In addition to $v$, the bare parameters of the Higgs potential 
are $m_h^2, m_H^2, m_A^2, m_{H^\pm}^2, \alpha, \beta, M^2, T_h^{}, T_H^{}$, 
where $T_h^{}$ and $T_H^{}$ are tadpoles of $h$ and $H$, respectively.
The tadpole parameters are fixed by the stationary condition 
at each order of perturbation. 
At the tree level, we set $T_h=T_H^{}=0$, 
while at one loop level $T_h$ and $T_H$ are chosen to make 
the renormalized one-point functions for $h$ and $H$ to be zero. 
They are expressed in terms of the Lagrangian parameters as 
\begin{eqnarray}
   T_H^{} =  T_1^{} \cos\alpha + T_2^{} \sin\alpha, \,\,\,\,
   T_h^{} =  - T_1^{} \sin\alpha + T_2^{} \cos\alpha, 
\end{eqnarray}
with 
\begin{eqnarray}
   T_1 &=& m_3^2 v_2 - m_1^2 v_1 
           - \frac{1}{2} \lambda_1 v_1^3 
           - \frac{1}{2} 
 (\lambda_3+\lambda_4+\lambda_5) v_1 v_2^2, \\
   T_2 &=& m_3^2 v_1 - m_2^2 v_2 
           - \frac{1}{2} \lambda_2 v_2^3 
           - \frac{1}{2} 
 (\lambda_3+\lambda_4+\lambda_5) v_1^2 v_2.
\end{eqnarray}
The renormalized parameters are defined by shifting 
the bare parameters as  
\begin{eqnarray}
 T_{h,H}^{} &\to& 0 + \delta T_{h,H}^{}, \\
 m_{\phi_i}^2 &\to& m_{\phi_i}^2 + \delta m_{\phi_i}^2, \\
 \alpha &\to& \alpha + \delta \alpha, \\
 \be &\to& \be + \delta \be, \\
 M^2 &\to& M^2 + \delta M^2,   
\end{eqnarray}
where $\phi_i$ represents $H$, $h$, $A$ and $H^\pm$. 
The introduction of the wave function renormalization factors 
for the Higgs bosons is rather complicated because the mixing 
between scalar bosons with the same quantum number 
should be taken into account. According to the method explained 
in Appendix~\ref{app:reno-mixing}, we define  
\begin{eqnarray}
\left[  \begin{array}{c} H \\ h \\
        \end{array}\right]
\to
\left[  \begin{array}{cc} 
          1 + \frac{1}{2} \delta Z_H^{} &  \delta \al + \delta C_h    \\
          - \delta \al + \delta C_h     & 1 + \frac{1}{2} \delta Z_h^{} \\
        \end{array}\right]
\left[  \begin{array}{c} H \\ h \\
        \end{array}\right], \label{mixwave1}
\end{eqnarray}
where 
$\delta Z_h$ ($\delta Z_H$) 
is the wave function factor of $h$ ($H$). Similarly, for 
the CP-odd scalar bosons and the charged scalar bosons we define 
\begin{eqnarray}
\left[  \begin{array}{c} z \\ A \\
        \end{array}\right]
\to
\left[  \begin{array}{cc} 
          1 + \frac{1}{2} \delta Z_z^{} &  \delta \be + \delta C_A    \\
          - \delta \be + \delta C_A     & 1 + \frac{1}{2} \delta Z_A^{} \\
        \end{array}\right]
\left[  \begin{array}{c} z \\ A \\
        \end{array}\right], \label{mixwave2}
\end{eqnarray}
and 
\begin{eqnarray}
\left[  \begin{array}{c} w^\pm \\ H^\pm \\
        \end{array}\right]
\to
\left[  \begin{array}{cc} 
          1 + \frac{1}{2} \delta Z_{w^\pm}^{} &  \delta \be + \delta C_{H^+}    \\
          - \delta \be + \delta C_{H^+}     & 1 + \frac{1}{2} \delta Z_{H^\pm}^{} \\
        \end{array}\right]
\left[  \begin{array}{c} w^\pm \\ H^\pm \\
        \end{array}\right], \label{mixwave3}
\end{eqnarray}
respectively, where 
$\delta Z_A$ $\delta Z_{H^\pm}^{}$ are the wave function 
renormalization factors for the physical CP-odd and charged 
Higgs bosons $A$ and $H^\pm$.  In addition,   
we introduced the ``wavefunction'' factors 
$\delta Z_z$ and $\delta Z_w$ for the Nambu-Goldstone bosons $z$ and
$w^\pm$, which are massless in the Landau gauge.
However, $\delta Z_z$ and $\delta Z_w$ will not be used 
in our calculation. 

There are sixteen counter-term parameters 
($\delta T_{h,H}$, 
$\delta m_{\phi_i}^2$,
$\delta Z_{\phi_i}$,
$\delta \alpha$,
$\delta \beta$,  
$\delta C_h$,
$\delta C_A$, 
$\delta C_{H^+}$ and $\delta M^2$), where 
$\phi_i=H,h,A$ and $H^\pm$, and 
$\delta C_h, \delta C_A$ and $\delta C_{H^+}$ 
are defined via Eqs.~(\ref{mixwave1})-(\ref{mixwave3}). 
The first fifteen of them 
are determined by imposing the renormalization condition 
to the one and two point functions. 
The conditions are shown below in order. 

The tadpole condition requires that the renormalized 
one-point functions for $h$ and $H$ must satisfy    
\begin{eqnarray}
  \Gamma_{h} = 0, \,\,\,\, \Gamma_{H}^{} = 0,   \label{rc_tad}
\end{eqnarray}
with $\Gamma_{h,H}=T^{\rm 1PI} + \delta T_{h,H}$. Thus,   
\begin{eqnarray}
  \delta T_{h}^{} = - T_h^{\rm 1PI},\,\,\,\,
  \delta T_{H}^{} = - T_H^{\rm 1PI}, 
\end{eqnarray}
where $T_{h,H}^{\rm 1PI}$ are the contributions 
of one-particle-irreducible (1PI) diagrams. 
The explicit expressions of the contributions 
to $T_{h,H}^{\rm 1PI}$ in the THDM are given in 
Appendix~\ref{app:1PI_THDM}. 

The relevant renormalized two-point functions for $h,H,A,H^\pm$ 
can be expressed as 
\begin{eqnarray}
 \Gamma_{hh}^{}(p^2) &=& \Pi_{hh}^{\rm 1PI}(p^2) +
   \left\{ (p^2-m_h^2) (1+\delta Z_h) - \delta m_h^2  
    +\frac{\sin^2\al}{\cos\be} \frac{\delta T_1}{v} 
    +\frac{\cos^2\al}{\sin\be} \frac{\delta T_2}{v} 
\right\},       \\
 \Gamma_{HH}^{}(p^2) &=&  \Pi_{HH}^{\rm 1PI}(p^2)+
 \left\{ (p^2-m_H^2) (1+\delta Z_H^{}) - \delta m_H^2  
    +\frac{\cos^2\al}{\cos\be} \frac{\delta T_1}{v} 
    +\frac{\sin^2\al}{\sin\be} \frac{\delta T_2}{v} 
\right\},       \\
 \Gamma_{AA}^{}(p^2) &=& \Pi_{AA}^{\rm 1PI}(p^2)+
   \left\{ \frac{}{}(p^2-m_A^2) (1+\delta Z_A^{}) - \delta m_A^2  
\right.
\nonumber\\
&&
\left.   
 +\left(\frac{\sin^2\be}{\cos\be}-\cos\be +\frac{1}{\cos\be} \right) 
                                                  \frac{\delta T_1}{2v} 
    +\left(\frac{\cos^2\be}{\sin\be}-\sin\be +\frac{1}{\sin\be} \right) 
                                                  \frac{\delta T_2}{2v} 
\right\},   \\
 \Gamma_{H^+H^-}^{}(p^2) &=& \Pi_{H^+H^-}^{\rm 1PI}(p^2)
+ 
   \left\{ \frac{}{}(p^2-m_{H^\pm}^2) (1+ \delta Z_{H^\pm}^{})
 - \delta m_{H^\pm}^2  
\right.
\nonumber\\
&&
\left.   
 +\left(\frac{\sin^2\be}{\cos\be}-\cos\be +\frac{1}{\cos\be} \right) 
                                                  \frac{\delta T_1}{2v} 
    +\left(\frac{\cos^2\be}{\sin\be}-\sin\be +\frac{1}{\sin\be} \right) 
                                                  \frac{\delta T_2}{2v} 
\right\},  
\end{eqnarray}
where $\Pi_{\phi\phi}^{\rm 1PI}(p^2)$ are the 1PI
diagram contributions to the self energies. 
Their expressions and those of $\delta T_1$ and $\delta T_2$ 
are summarized in Appendix~\ref{app:1PI_THDM}. 

By imposing the on-shell conditions  
\begin{eqnarray}
 {\rm Re} \Gamma_{\phi_i\phi_i}(m_{\phi_i}^2) &=& 0, \,\,\,\, 
  \left. \hspace*{1cm} \frac{\partial}{\partial p^2}
 {\rm Re} 
 \Gamma_{\phi_i\phi_i}(p^2) \right|_{p^2=m_{\phi_i}^2} = 1,  \label{rc_mass}
\end{eqnarray}
where $\phi_i$ represents $h$,$H$,$A$ and $H^\pm$, we determine 
$\delta m_h^2$,
$\delta m_H^2$,
$\delta m_A^2$,
$\delta m_{H^\pm}^2$,
$\delta Z_h^{}$,
$\delta Z_H^{}$,
$\delta Z_A^{}$ and
$\delta Z_{H^\pm}^{}$. 

    The condition that there is no mixing between 
    CP-even scalar bosons $h$ and $H$ on each mass shell; i.e., 
\begin{eqnarray}
  \Gamma_{hH}^{}(m_h^2) = 0, \,\,\,\,  
  \Gamma_{hH}^{}(m_H^2) = 0,      \label{rc_cpemix}
\end{eqnarray}
    determines $\delta \alpha$ and $\delta C_h$, where 
\begin{eqnarray}
 \Gamma_{hH}^{}(p^2) &=& \tilde{\Pi}_{hH}^{}(p^2) + 
    (2 p^2-m_h^2 - m_H^2) \delta C_h 
          - (m_H^2 - m_h^2) \delta \alpha, 
\end{eqnarray}
with
\begin{eqnarray}
 \tilde{\Pi}_{hH}^{}(p^2) &=& \Pi_{hH}^{\rm 1PI}(p^2) 
    + \cos\al\sin\al \left(
    - \frac{1}{\cos\be} \frac{\delta T_1}{v} 
    +\frac{1}{\sin\be} \frac{\delta T_2}{v} 
 \right).
\end{eqnarray}
Hence, we obtain 
\begin{eqnarray}
 \delta \alpha &=& + \frac{1}{2} \frac{1}{m_H^2-m_h^2} 
                   \left\{ \tilde{\Pi}_{hH}(m_H^2) 
                         + \tilde{\Pi}_{hH}(m_h^2)  \right\}, \\
 \delta C_h &=& - \frac{1}{2} \frac{1}{m_H^2-m_h^2} 
                   \left\{ \tilde{\Pi}_{hH}(m_H^2) 
                         - \tilde{\Pi}_{hH}(m_h^2)  \right\}.
\end{eqnarray}
The expression of the 1PI diagrams 
$\tilde{\Pi}_{hH}^{}(p^2)$ 
can be obtained from those of $\Pi_{hH}^{}$, $\delta T_1$ and $\delta T_2$
in Appendix~\ref{app:1PI_THDM}.

The parameter $\delta\beta$, $\delta C_A$ and $\delta C_{H^+}$ are 
determined by the conditions for the two-point functions
of $z$-$A$ and $w^\pm$-$H^\pm$ mixings. 
For the CP-odd sector, we require 
\begin{eqnarray}
 \Gamma_{zA}^{}(0) &=& 0, \label{eq:cond-zA1}\\
  \Gamma_{zA}^{}(m_A^2) &=& 0,     \label{eq:cond-zA2}
\end{eqnarray}
where 
\begin{eqnarray}
 \Gamma_{zA}^{}(p^2) &=& \tilde{\Pi}_{zA}^{}(p^2) + 
  (2 p^2-m_A^2) \delta C_A  + m_A^2 \delta \beta,
\end{eqnarray}
with
\begin{eqnarray}
 \tilde{\Pi}_{zA}^{}(p^2) &=& \Pi_{zA}^{\rm 1PI}(p^2)  
    - \sin\be \frac{\delta T_1}{v} 
    + \cos\be \frac{\delta T_2}{v} .
\end{eqnarray}
Due to the Nambu-Goldstone's theorem,  
$\tilde{\Pi}_{zA}^{}(0)=0$ is ensured, 
so that we obtain, 
from Eqs.~(\ref{eq:cond-zA1}) and (\ref{eq:cond-zA2}), 
\begin{eqnarray}
 \delta C_A = \delta \beta = -\frac{1}{2 m_A^2} \tilde{\Pi}_{zA}(m_A^2). 
\end{eqnarray}
For the charged sector, from the condition 
\begin{eqnarray}
 \Gamma_{w^\pm H^\mp}^{}(0) &=& 0, \label{eq:cond-wH1}
\end{eqnarray}
where 
\begin{eqnarray}
 \Gamma_{w^\pm H^\mp}^{}(p^2) &=& 
 \tilde{\Pi}_{w^\pm H^\mp}^{}(p^2) + 
 (2 p^2-m_{H^\pm}^2) \delta C_{H^+}  + 
 m_{H^\pm}^2 \delta \beta,
\end{eqnarray}
with
\begin{eqnarray}
 \tilde{\Pi}_{w^\pm H^\mp}^{}(p^2) &=& 
 \tilde{\Pi}_{w^\pm H^\mp}^{\rm 1PI}(p^2)  
    - \sin\be \frac{\delta T_1}{v} 
    + \cos\be \frac{\delta T_2}{v} ,
\end{eqnarray}
we obtain 
\begin{eqnarray}
   \delta C_{H^+} = \delta \beta.
\end{eqnarray}

We note that due to the Ward-Takahashi identity, the condition (\ref{eq:cond-zA2}) 
is equivalent to the following condition on the mixing 
between the gauge boson and the Higgs boson: 
\begin{eqnarray}
  \Gamma_{ZA}^{}(m_A^2) = 0,  \label{eq:cond_ZA}
\end{eqnarray}
where the two-point function of $ZA$ is written as 
\begin{eqnarray}
  \Gamma_{ZA}^{\mu}(p^2) = - i p^\mu   \Gamma_{ZA}^{}(p^2), 
\end{eqnarray}
and the form factor $\Gamma_{ZA}^{}(p^2)$ 
is expressed as  
\begin{eqnarray}
  \Gamma_{ZA}^{}(p^2) &=& (\delta \be + \delta C_A) m_Z^{} 
                     + \Gamma_{ZA}^{\rm 1PI}(p^2).
\end{eqnarray}
In the above equation,
the counter term parameters $(\delta \be + \delta C_A)$ 
comes from the Higgs kinematic terms of the Lagrangian 
as the consequence of the shift of the parameters:
\begin{eqnarray}
 {\cal L} &=& m_Z^{} (\partial_\mu z) Z^\mu \nonumber\\
&\to&  + (\delta \be + \delta C_A) m_Z^{}  (\partial_\mu A) Z^\mu 
+ \cdot\cdot\cdot. 
\end{eqnarray}
With the expressions of $\Gamma_{ZA}^{\rm 1PI}(p^2)$ and 
$\Gamma_{zA}^{\rm 1PI}(p^2)$ presented in
Appendix~\ref{app:1PI_THDM}, 
one can explicitly check the equivalence of the conditions 
of Eqs.~(\ref{eq:cond-zA2}) and (\ref{eq:cond_ZA}).
Similarly, instead of the condition (\ref{eq:cond-zA2}), 
the alternative condition 
\begin{eqnarray}
 \Gamma_{w^\pm H^\mp}^{}(m_{H^\pm}^{2}) &=& 0 \label{eq:cond-wH2}
\end{eqnarray}
may be used to determine $\delta \beta$. 
In this case, $\delta \beta$ (let us denote it as $\delta \beta'$) 
is given by 
\begin{eqnarray}
 \delta \beta' (=\delta C_{H^+}' = \delta C_A') 
= -\frac{1}{2 m_{H^\pm}^2} \tilde{\Pi}_{w^\pm H^\mp}(m_{H^\pm}^2). 
\end{eqnarray}
It is easy to check that 
the difference between $\delta \beta$ and $\delta \beta'$ is finite.
This finite difference is due to different choice of renormalization
prescription in loop calculations.
In this paper, we adopt $\delta \beta$ determined 
from Eqs.~(\ref{eq:cond-zA1}) and (\ref{eq:cond-zA2}). 

We have determined all the renormalization parameters 
but $\delta M^2$ of the Higgs sector
by applying the renormalization conditions to various 
one- and two-point functions. 
However, the renormalization of $M^2$ has to be 
discussed in the context of three-point functions. 
Below, we consider the renormalization calculation 
for the three-point $hZZ$ and $hhh$ vertices.

The tree level $hZZ$ coupling can be read out 
from the kinematic term of the Lagrangian:
\begin{eqnarray}
{\cal L}_{hZZ}^{} &=&
+ \frac{m_Z^2}{v} \sin(\be-\al) g_{\mu\nu}^{} Z^\mu Z^\nu h     
+ \frac{m_Z^2}{v} \cos(\be-\al) g_{\mu\nu}^{} Z^\mu Z^\nu H.   
\end{eqnarray} 
In terms of the general form factors of the $hZZ$ coupling, 
cf. Eq.~(\ref{eq:hzz-form}), we get 
\begin{eqnarray}
 M_1^{hZZ{\rm (tree)}} &=& \frac{2 m_Z^2}{v} \sin(\be-\al),
\,\,\,\, M_2^{hZZ{\rm (tree)}} = M_3^{hZZ{\rm (tree)}} = 0. 
\label{treehzz_THDM} 
\end{eqnarray} 
On the other hand, the tree level coupling constants of $hhh$ and $hhH$ 
are given from the Higgs potential. By 
using the mass relations of Eqs.~(\ref{mass_THDM1}) -
(\ref{mass_THDM5}), 
each coupling constant can be expressed  
in terms of Higgs boson masses and mixing angles: 
\begin{eqnarray}
\!\!\!\!\!\!
\!\!\!\!\!\!
\lambda_{hhh}^{} &=& 
\frac{-1}{4v\sin 2\be} 
\left[
\left\{\frac{}{} \cos(3\al-\be) + 3 \cos(\al+\be) \right\} m_h^2 
- 4 \cos^2(\al-\be) \cos(\al+\be) M^2
\right], \\
\!\!\!\!\!\!
\!\!\!\!\!\! 
\lambda_{hhH}^{} &=& 
\frac{-1}{2v\sin 2\be} 
\left\{\frac{}{}
 \cos(\al-\be) \sin 2\al (2 m_h^2 + m_H^2)
- \cos(\al-\be) \left(3 \sin 2\al -\sin 2\be \right)  M^2
\right\} , 
\end{eqnarray}
The tree level form factor for the $hhh$ coupling, 
$\Gamma_{hhh}^{\rm (tree)}$, is thus given by 
\begin{eqnarray}
\Gamma_{hhh}^{\rm (tree)} &=& 3 ! \lambda_{hhh}^{}. 
\end{eqnarray}
Note that in the SM-like limit ($\al=\be-\pi/2$), 
the form factors of $hZZ$ and $hhh$ couplings 
take the same form as in the SM: 
\begin{eqnarray}
 M_1^{hZZ {\rm (tree)}} &=& \frac{2 m_Z^2}{v}, 
 \,\,\,\, M_2^{hZZ {\rm (tree)}} = M_3^{hZZ {\rm (tree)}} = 0, 
\end{eqnarray} 
\begin{eqnarray}
\Gamma_{hhh}^{\rm (tree)} &=& - \frac{3 m_h^2}{v}, 
\end{eqnarray}
while the heavier Higgs boson $H$ does not couple to the gauge bosons 
and also $\lambda_{hhH}^{} = 0$.

Now, we discuss the renormalized vertices of 
$hZZ$ and $hhh$. From the kinematic term of the Higgs sector, 
we obtain the counter terms to the form factors of the $hZZ$ vertex 
as follows. 
\begin{eqnarray}
{\cal L}_{hZZ}^{} &=& 
-\frac{m_Z^2}{v} \sin(\al-\be) g_{\mu\nu} Z^\mu Z^\nu h
+\frac{m_Z^2}{v} \cos(\al-\be) g_{\mu\nu} Z^\mu Z^\nu H \nonumber \\
&\to& 
- \frac{m_Z^2}{v} 
\left\{
\sin(\al-\be) \left( 
1 + \frac{\delta m_Z^2}{m_Z^2} - \frac{\delta v}{v} 
+ \delta Z_Z^{} + \frac{1}{2} \delta Z_h 
\right) \right.
\nonumber \\
&& \left. \frac{}{} 
+ \cos(\al-\be) (-\delta \be - \delta C_h)
\right\} g_{\mu\nu}^{}  Z^\mu Z^\nu h + \cdot\cdot\cdot .
\end{eqnarray} 
Thus, we obtain the counter terms for the $hZZ$ form factors as
\begin{eqnarray}
 \delta M_1^{hZZ} &=& 
- \frac{2 m_Z^2}{v} 
\left\{
\sin(\al-\be) \left( 
 \frac{\delta m_Z^2}{m_Z^2} - \frac{\delta v}{v} 
+ \delta Z_Z^{} + \frac{1}{2} \delta Z_h 
\right) \right.
\nonumber \\
&& \left. \frac{}{} 
+ \cos(\al-\be) (-\delta \be - \delta C_h)
\right\},   \\
\delta M_2^{hZZ} &=& \delta M_3^{hZZ} = 0.
\end{eqnarray} 

The counter term for the $hhh$ vertex is obtained 
from the shifting the bare $hhh$ coupling $\delta \lambda_{hhh}^{}$ 
and the wave function and mixing renormalization 
of the $hhh$ and $hhH$ vertices as
\begin{eqnarray}
\lambda_{hhh}^{} &\to& \lambda_{hhh}^{} + \delta \lambda_{hhh}^{}, \\
hhh &\to& 
\left\{ 
\left( 1 + \frac{1}{2} \delta Z_h \right) h 
                + \cdot\cdot\cdot \right\}^3 \to   
\left( 1 + \frac{3}{2} \delta Z_h \right) h^3 + \cdot\cdot\cdot, \\
hhH &\to& 
\left\{ \left( 1 + \frac{1}{2} \delta Z_h \right) h 
        + \cdot\cdot\cdot 
\right\}^2 
\left\{
\left( \delta \al + \delta C_h \right) h  
+ \cdot\cdot\cdot \frac{}{}\right\} 
\to
\left( \delta \al + \delta C_h \right) h^3 
+ \cdot\cdot\cdot .
\end{eqnarray}
Thus, the counter term for the $\Gamma_{hhh}^{}$ is obtained as  
\begin{eqnarray}
\delta \Gamma_{hhh}^{} 
&=& 3 ! \left\{ \delta \lambda_{hhh}^{} 
+ \frac{3}{2}  \lambda_{hhh}^{}  \delta Z_h 
+ \lambda_{hhH}^{}  \left( \delta \al + \delta C_h \right) 
\right\} \\
&=& 3 !\lambda_{hhh}^{} \left(\frac{3}{2} \delta Z_h^{} - \frac{\delta v}{v} 
\right)
 + 3 ! \lambda_{hhH}^{} \delta C_h 
+ C_1 \delta m_h^2 + C_2 \delta \al + C_3 \delta \beta + C_4 \delta M^2,
\end{eqnarray}
where 
\begin{eqnarray}
C_1 &=& \frac{-1}{4 v \sin 2\be} 
   \left\{ \cos(3\al-\be) + 3 \cos(\al+\be)\right\}, \\
C_2 &=& \frac{-1}{2 v \sin 2\be} 
    \cos(\al-\be) \sin 2\al (m_H^2-m_h^2), \\
C_3 &=& \frac{\cos(\al-\be)}{4 v \sin^2 2\be} 
   \left[ 
\left\{ \frac{}{}
4+ \cos 2(\al-\be) + 3 \cos 2(\al+\be)\right\} m_h^2 
\right.\nonumber\\
&& -
\left. \left\{5+ \cos 2(\al-\be) - \cos 4\be + 3 \cos 2(\al+\be)
\frac{}{}\right\} M^2
\right], \\
C_4 &=& \frac{1}{v \sin 2\be} 
   \cos^2(\al-\be) \cos(\al+\be).
\end{eqnarray}

Up to now we have not explicitly discussed the renormalization 
condition to determine the counter term of the soft-breaking mass, 
$\delta M^2$. We chose to fix this parameter 
in the minimal subtraction method. 
Namely, we require the condition that 
the remaining divergent term proportional to 
$\Delta=1/\epsilon + \ln\mu^2$ 
in the $hhh$ vertex is canceled by the counter term $\delta M^2$. 
In the present model, $\delta M^2/M^2$ is found to be 
\begin{eqnarray}
\frac{\delta M^2}{M^2} 
&=&\frac{1}{16 \pi^2 v^2} 
 \left\{ \frac{}{}
     2 N_c ( m_t^2 \cot^2 \be + m_b^2 \tan^2 \be ) 
\right.\nonumber\\
&& \hspace*{2cm}
\left.
+ 4 M^2 - 
2 m_{H^\pm}^2 - m_A^2 
+ \frac{\sin 2\al}{\sin 2\be} (m_H^2 - m_h^2 ) 
\right\}    \Delta, \label{rc_M}
\end{eqnarray}
where $\Delta=1/\epsilon+\ln \mu^2$ with $D=4-2\epsilon$.

Finally, the renormalized form factors for $hZZ$ and $hhh$ 
couplings are calculated by 
\begin{eqnarray}
 M_i^{hZZ}(p_1^2,p_2^2,q^2)
 &=&  M_i^{hZZ(\rm tree)} + 
                M_i^{hZZ(\rm 1PI)}(p_1^2,p_2^2,q^2) 
 + \delta M_i^{hZZ},  (i=1-3), \label{reno_hZZ}\\
 \Gamma_{hhh}^{}(p_1^2,p_2^2,q^2)
 &=& \Gamma_{hhh}^{\rm tree} 
                +\Gamma_{hhh}^{\rm 1PI}(p_1^2,p_2^2,q^2) 
+ \delta \Gamma_{hhh}^{},  \label{reno_hhh}
\end{eqnarray}
where the momentum $q^\mu$ in Eq.~(\ref{reno_hZZ}) 
is that of the external Higgs boson line, and 
all the counter terms are completely determined 
by the renormalization conditions 
in Eqs.~(\ref{rc_tad}), (\ref{rc_mass}), 
(\ref{rc_cpemix}), (\ref{eq:cond-zA1}), (\ref{eq:cond-zA2}), 
(\ref{eq:cond-wH1}) and (\ref{rc_M}).
All the explicit results for the 1PI diagrams which contribute 
to the form factors are summarized in Appendix~\ref{app:1PI_THDM}.

\section{Large mass expression in the SM-like regime}\label{sec:5}

The renormalized coupling constants $hZZ$ and $hhh$ are 
evaluated by the formulae given in Eqs.~(\ref{reno_hZZ}) 
and (\ref{reno_hhh}). 
The deviation from the SM predictions can occur due to two 
sources: the mixing effect which appears
in the tree level, and the quantum correction effect due to 
the loop contribution of the extra Higgs bosons. 
If the mixing between the two CP-even Higgs bosons is large  
(e.g., $\sin(\al-\be)^2 \sim 0.3-0.7$),
the $hZZ$ form factor $M_1^{hZZ}$ in the THDM significantly
differs from the SM prediction
already at tree level by the factor of $\sin(\be-\al)$; 
cf. Eq.~(\ref{treehzz_THDM}). 
In such a case, we may be able to obtain an indirect but explicit 
evidence of extended Higgs sectors at the LHC or 
at the early stage of the LC experiments.

On the other hand, 
the $hZZ$ coupling may be close to the SM prediction;  
i.e., in the SM-like regime\cite{SMlikeTHDM} where  
$\sin^2(\alpha-\beta) \simeq 1$.
Define $x=\beta-\alpha-\pi/2$. As $x \ll 1$, 
the tree level $hZZ$ and $hhh$ couplings can be expressed as follows:
\begin{eqnarray}
 M_1^{hZZ{\rm (tree)}} &=&  \frac{2 m_Z^2}{v}
      \left\{ 1  - \frac{1}{2} x^2
        + {\cal O}\left(x^4\right)\right\}, \label{xdephzz}    \\
 \Gamma_{hhh}^{\rm (tree)} &=& - \frac{3 m_h^2}{v}
      \left\{ 1  + \frac{3}{2} 
                   \left(1 - \frac{4 M^2}{3 m_h^2}\right) x^2
        + {\cal O}\left(x^3\right)
      \right\}. \label{xdephhh}    
\end{eqnarray}
In the limit of $x \to 0$, 
the mixing effect vanishes and 
the $hZZ$ and $hhh$ couplings coincide with the SM formulas. 
From Eq.~(\ref{xdephhh}), we find that 
in the SM-like regime the $hhh$ coupling constant is reduced 
from the SM value as long as $M^2 > 3 m_h^2/4 $. 
Keeping only the leading one loop contributions of the heavier Higgs bosons 
and the top quark, 
the expressions for the one loop corrected $hZZ$ and $hhh$ 
couplings are given in the SM-like regime ($x \ll 1$) as   
\begin{eqnarray}
 M_1^{hZZ} &=&  \frac{2 m_Z^2}{v}
      \left\{ 1  - \frac{1}{2} x^2
              + \frac{1}{64 \pi^2 v^2} 
                 (m_H^2 + m_A^2 + 2 m_{H^\pm}^2) 
              - \frac{m_{H}^2}{96 \pi^2 v^2} 
                         \left(1 - \frac{M^2}{m_H^2}\right)^2 \right.\nn\\
&&\left.  \!\!\!\!\!\!\!\!\!\!
              - \frac{m_{A}^2}{96 \pi^2 v^2} 
                         \left(1 - \frac{M^2}{m_A^2}\right)^2 
              - \frac{m_{H^\pm}^2}{48 \pi^2 v^2} 
                         \left(1 - \frac{M^2}{m_{H^\pm}^2}\right)^2
              - \frac{5 N_{c_t} m_t^2}{96 \pi^2 v^2} 
        + {\cal O}\left(x^4, \frac{p^2_i}{v^2},
                             \frac{m_h^2}{v^2} \right)
               \right\},  \label{m2THDM}    \\
 \Gamma_{hhh}^{} &=& - \frac{3 m_h^2}{v}
      \left\{ 1  + \frac{3}{2} 
                   \left(1 - \frac{4 M^2}{3 m_h^2}\right) x^2
              + \frac{m_{H}^4}{12 \pi^2 m_h^2 v^2} 
                         \left(1 - \frac{M^2}{m_H^2}\right)^3 
              + \frac{m_{A}^4}{12 \pi^2 m_h^2 v^2} 
                         \left(1 - \frac{M^2}{m_A^2}\right)^3 \right.\nn\\
  &&\left.
              + \frac{m_{H^\pm}^4}{6 \pi^2 m_h^2 v^2} 
                         \left(1 - \frac{M^2}{m_{H^\pm}^2}\right)^3
              - \frac{N_{c_t} m_t^4}{3 \pi^2 m_h^2 v^2} 
        + {\cal O}\left(x^3, \frac{p^2_i m_\Phi^2}{v^2 m_h^2},
                             \frac{m_\Phi^2}{v^2}, 
                             \frac{p^2_i m_t^2}{v^2 m_h^2},
                             \frac{m_t^2}{v^2} \right)
      \right\}, \label{m4THDM}    
\end{eqnarray}
where $m_\Phi^{}$ represents the masses of the 
heavier Higgs bosons $H$, $A$ and $H^\pm$.\footnote{
Although the expression in Eq.~(\ref{m4THDM}) does not depend on
$\tan\beta$, the allowed value of $\tan\beta$ is constrained to be
${\cal O}(1)$ due to the requirement of 
the perturbative unitarity when large values of $m_\Phi^{}$
are taken with $M=0$.
Hence, the large deviation from the SM prediction occurs  
at $\tan\beta = {\cal O}(1)$. 
We note that the parameter set 
$m_H^{}=m_A^{}=m_{H^\pm}^{}$, $M=0$, $\alpha=\beta-\pi/2$ and 
$\tan\beta=1$ 
corresponds to $\lambda_1=\lambda_2=\lambda_3=(m_h^2+m_H^2)/v^2$ and 
$\lambda_4=\lambda_5=-m_H^2/v^2$ at the tree level.  
}
As expected, there are quartic power terms of the 
heavier Higgs boson masses in $\Gamma_{hhh}^{}$. 
The difference from the top-mass contribution is the 
suppression factor of $(1-M^2/m_{\Phi}^2)^3$ and the sign.  
The large correction occurs in the case of small $M^2$. 
The largest corresponds to the limit of $M^2 \to 0$. 
In this case, a greater positive deviation from the SM prediction 
is obtained for a larger $m_{\Phi}$.  
However, since $m_{\Phi}^2$ is originated 
from the electroweak symmetry breaking and is proportional to
$\lambda_i$, 
a too large value of $m_{\Phi}^2$ is forbidden by the requirement of 
the perturbative unitarity\cite{LQT}.  
Furthermore, 
the direction of the deviation 
induced by the heavy Higgs boson (bosonic) loops 
is opposite to that by the top quark (fermionic) loops.

\section{Numerical Evaluation}\label{sec:6}

\begin{figure}[t]
\vspace*{-12mm}
\begin{center}
\hspace*{-3mm}
\includegraphics[width=10cm,height=8cm]{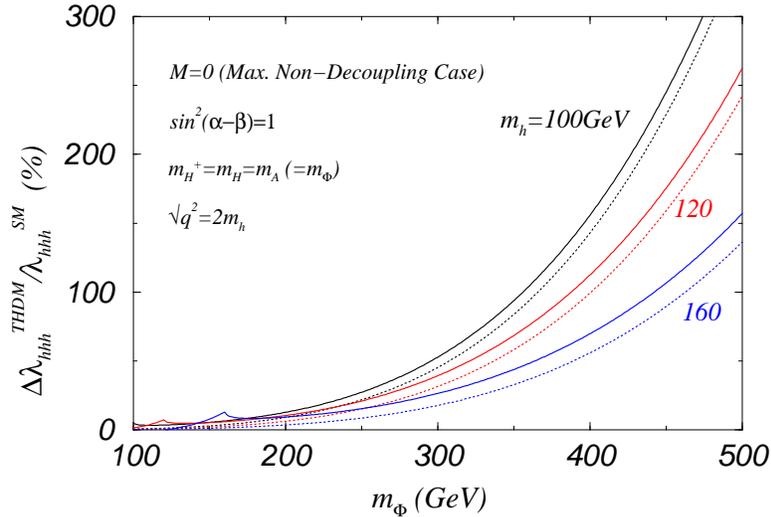}
\end{center}
\vspace*{-8mm}
\caption{ 
              ($\Delta\lambda_{hhh}^{THDM}/\lambda_{hhh}^{SM}$) is shown
              as a function of  $m_{\Phi} 
              (\equiv m_H^{}=m_A^{}=m_{H^\pm}^{})$.  
              The results of the full one loop calculation 
              are shown as solid curves, while the quartic mass
              ($m^4_{\Phi}$) 
              contributions, given in 
              Eq.~(\ref{m4THDM}), are plotted as dotted curves.
}\label{fig:newletfig1}
\end{figure}

In this section, we present results of our numerical evaluation 
for the effective 
$hZZ$ and $hhh$ couplings predicted by the SM and the THDM  
at the one loop order. 
We define the deviation from the SM prediction by  
\begin{eqnarray}
   \left( \frac{\Delta g_{hZZ}^{THDM}}{g_{hZZ}^{SM}}\right) (q^2)
    &=&\frac{M^{hZZ(THDM)}_1(q^2,m_Z^2,m_h^2) 
         - M^{hZZ(SM)}_1(q^2,m_Z^2,m_h^2) }
                 {M^{hZZ(SM)}_1(q^2,m_Z^2,m_h^2) },\\
 \left(  \frac{\Delta \lambda_{hhh}^{THDM}}{\lambda_{hhh}^{SM}}\right) (q^2) 
     &=&\frac{\Gamma_{hhh}^{THDM}(m_h^2,m_h^2,q^2) 
 - \Gamma_{hhh}^{SM}(m_h^2,m_h^2,q^2) }{\Gamma_{hhh}^{SM}(m_h^2,m_h^2,q^2) },
\end{eqnarray}
where the SM form factors 
$M^{hZZ(SM)}_1(p_1^2,p_2^2,q^2)$ and 
$\Gamma_{hhh}^{SM}(p_1^2,p_2^2,q^2)$ 
are evaluated by using Eqs.~(\ref{sm_hzz}) and (\ref{sm_hhh}),
and those of the THDM,  
$M^{hZZ(THDM)}_1(p_1^2,p_2^2,q^2)$ and 
$\Gamma_{hhh}^{SM}(p_1^2,p_2^2,q^2)$, 
are given by Eqs.~(\ref{reno_hZZ}) and (\ref{reno_hhh}).
In the following numerical analysis, 
we fix $\sqrt{q^2} = 2 m_h$ except for Fig.~4. 
We show the momentum dependence of the deviation in the $hhh$ 
form factor in Fig.~4.  
Throughout this section, 
the mass of the top quark is set to be $m_t = 175$ GeV.

\begin{figure}[t]
\begin{center}
\hspace*{-3mm}
\includegraphics[width=10cm,height=8cm]{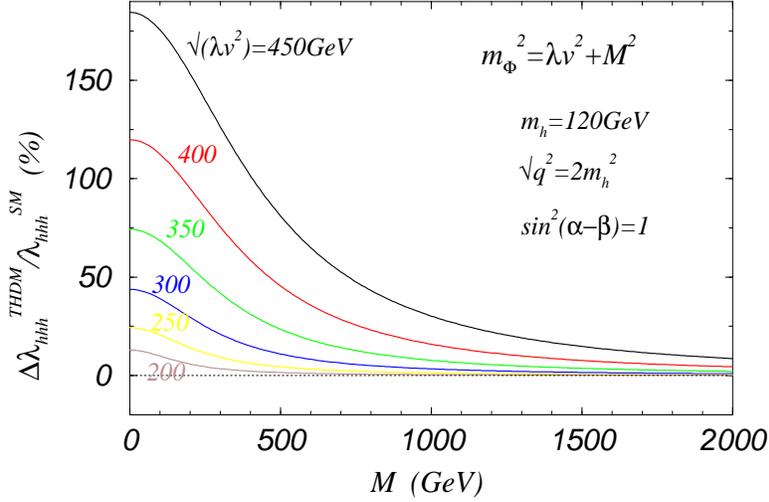}
\end{center}
\vspace*{-8mm}
\caption{
     The decoupling behavior of   
($\Delta\lambda_{hhh}^{THDM}/\lambda_{hhh}^{SM}$)
is shown.     The mass of the heavy Higgs bosons 
              $m_{\Phi} (\equiv m_H^{}=m_A^{}=m_{H^\pm}^{})$ 
              is given by $m_{\Phi}^2=\lambda v^2+M^2$.         }
\label{fig:letfig2}
\end{figure}

\subsection{The SM-like  limit}

First, we show the results in the SM-like limit 
($\sin^2(\alpha-\beta)=1$  or $x \to 0$ 
in Eqs.~(\ref{m2THDM}) and (\ref{m4THDM})), 
where the tree level expressions are coincide with the 
SM ones; cf. Eqs.~(\ref{xdephzz}) and (\ref{xdephhh}).   
In this case, the contributions to 
$\Delta g_{hZZ}^{THDM}$ and 
$\Delta \lambda_{hhh}^{THDM}$ 
only comes from radiative corrections.  
The form factor $M_1^{hZZ}$ receives the one loop effect of 
${\cal O}[m_\Phi^2/(16 \pi^2 v^2)]$ 
due to the heavy Higgs boson $\Phi^{}$ ($\Phi = H$, $A$ and $H^\pm$) 
with the suppression factor $(1 - M^2/m_\Phi^2)^2$. 
When $M=0$, where the non-decoupling loop effect is
maximal, the magnitude of the deviation 
$\Delta g_{hZZ}^{THDM}/g_{hZZ}^{SM}$  
becomes typically at most ${\cal O}(1)$ \%. 
On the other hand, the loop effect for the $hhh$ coupling is  
${\cal O}[m_\Phi^4/(16 \pi^2 v^2 m_h^2)]$ 
with the suppression factor $(1 - M^2/m_\Phi^2)^3$. 
The magnitude is larger than that for the $hZZ$ coupling 
by the enhancement factor of $m_\Phi^2/m_h^2$ \cite{lcws02kkosy,kkosy}. 
In Fig.~\ref{fig:newletfig1}, the one loop contribution of 
the heavy Higgs bosons to the $hhh$ coupling 
is shown for $m_h=100$, $120$ and $160$ GeV 
as a function of $m_\Phi^{}$, where
$m_\Phi^{} \equiv m_H^{} = m_A^{} = m_{H^\pm}^{}$, 
by assuming $\sin^2(\al-\be)=1$ and  $M^2=0$.
The deviation increases rapidly for large $m_\Phi$ values 
due to the quartic power dependence of $m_\Phi^{}$,  
and it amounts to 50 (100) \% for $m_\Phi^{}=300$ ($400$) GeV 
for $m_h=120$ GeV. 
The larger deviation is obtained for the smaller value of $m_h^{}$.
The small ``peak'' structure in Fig.~\ref{fig:newletfig1} originates from
the threshold contribution when $m_\Phi = m_h$ for $\sqrt{q^2}=2 m_h$,
where  $\sqrt{q^2}$ is the invariant mass of the virtual $h$, i.e. the
invariant mass of the two on-shell $h$ Higgs bosons in the $hhh$ vertex.

For non-zero values of $M$ ($0 < M^2 < m_\Phi^2$), the magnitude of 
the loop correction is suppressed by the factor $(1 - M^2/m_\Phi^2)^3$, 
and the non-decoupling effect vanishes when $M \simeq m_\Phi^{}$. 
In Fig.~3, we show the decoupling behavior of the heavier Higgs 
contribution
as a function of $M$ with fixed $\sqrt{\lambda v^2}=200-450$ GeV, in the
case of $\sin^2(\al-\be)=1$ and $m_h^{}=120$ GeV, where
the mass of the heavier Higgs bosons
$m_{\Phi}^{}$ ($=m_A^{}=m_H^{}=m_{H^\pm}$)
is given by $m_\Phi^2=\lambda v^2 + M^2$. 
(We note that $\lambda$ corresponds to 
 $\lambda_1 \cos^2 \beta + \lambda_2 \sin^2\beta - m_h^2/v^2 
  = \lambda_3 - m_h^2/v^2 = 
  - \lambda_4 = - \lambda_5$ in this case.)
As shown, the heavier Higgs boson contributions reduce rapidly for a larger
value of $M$. Nevertheless, 
a few tens of percent of the correction remains at
$M=1000$ GeV.

\begin{figure}[t]
\vspace*{-12mm}
\begin{center}
\hspace*{-3mm}
\includegraphics[width=10cm,height=8cm]{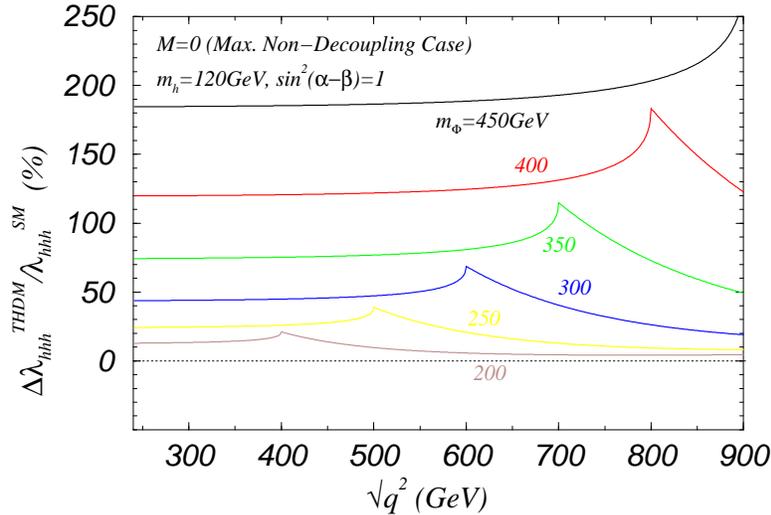}
\end{center}
\vspace*{-8mm}
\caption{     The momentum dependence of 
         ($\Delta\lambda_{hhh}^{THDM}/\lambda_{hhh}^{SM}$)
          is shown, where  
          $\sqrt{q^2}$ is the invariant mass of $h^\ast$ in 
          $h^\ast \to hh$, for each value
          of $m_\Phi^{}$ ($\equiv m_H^{}=m_A^{}=m_{H^\pm}^{}$) 
          when $m_h=120$ GeV, $\sin(\alpha-\beta)=-1$ and $M=0$.}
\end{figure}

In Fig.~4, we show the momentum dependence of the deviation 
in the effective $hhh$ coupling, 
$\Gamma_{hhh}(q^2)$ ($\equiv \Gamma_{hhh}(m_h^2, m_h^2,q^2)$), 
from the SM result as a function of the invariant mass
($\sqrt{q^2}$) of the virtual $h$ boson, for
various values of $m_\Phi^{}$ ($=m_A^{}=m_H^{}=m_{H^\pm}$)
with $\sin^2(\al-\be)=1$ and $m_h=120$ GeV.
Again, to show the maximal non-decoupling effect,
we have set $M$ to be zero.
The Higgs boson one loop contribution is always positive. 
Below the peak of the threshold of the heavy Higgs pair production, 
$\Gamma_{hhh}(q^2)$ is insensitive to $\sqrt{q^2}$. 
We note that the low $\sqrt{q^2}$  (but $\sqrt{q^2} \gsim 2 m_h$) 
is the most important region in the extraction of the $hhh$ 
coupling from the data of the double Higgs production mechanism, 
because the $h^\ast$ propagator $1/(q^2-m_h^2)$ in the signal 
process becomes larger.  
On the contrary, as we have shown in Fig.~\ref{fig:newSM2.eps} in 
Sec.~\ref{sec:2},  
the fermionic (top-quark) loop effect strongly depends on $\sqrt{q^2}$  
because of the threshold enhancement 
at $\sqrt{q^2}=2 m_t^{}$.

\begin{figure}[t]
\begin{center}
\hspace*{-3mm}
\includegraphics[width=8cm,height=6cm]{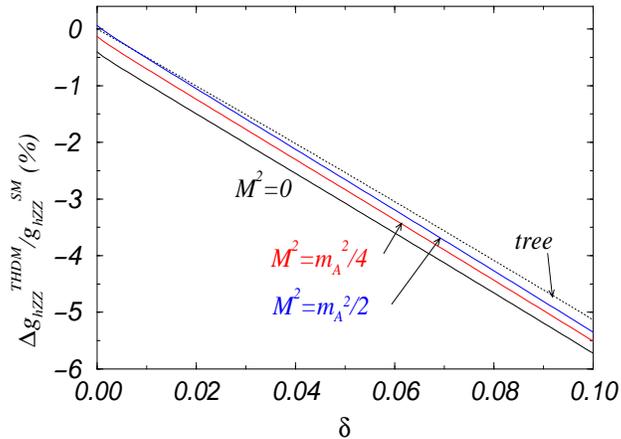}
\end{center}
\vspace*{-8mm}
\caption{
Deviation of the one loop renormalized (solid curves) and tree level
(dotted curves)
form factor $M_1^{hZZ}$ from the SM value is shown as a function 
of $\delta=\cos^2(\al-\be)$ for various $M$ values. 
The other parameters are set to be 
$m_h^{}=120$ GeV, $\tan\beta=2$, and 
$m_H^{}=m_{H^\pm}^{}=m_A^{}=300$ GeV.
}
\label{fig:zzh_tree_loop.eps}
\end{figure}

\subsection{The mixing angle dependence}

\begin{figure}[t]
\begin{center}
\hspace*{-3mm}
\includegraphics[width=8cm,height=6cm]{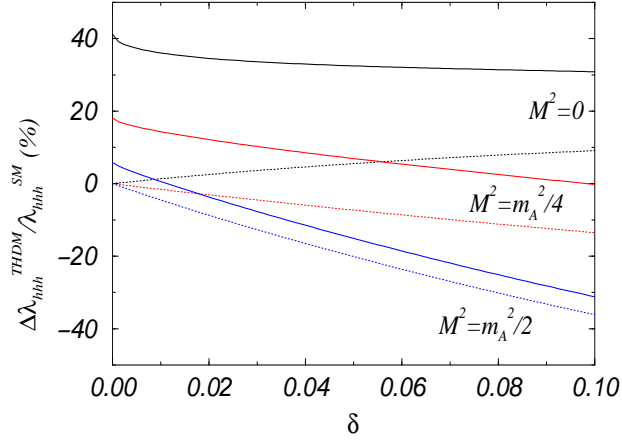}
\end{center}
\vspace*{-8mm}
\caption{
Deviation of the one loop renormalized (solid curves) and tree level
(dotted curves) $hhh$ form factor 
from the SM value is shown as a function 
of $\delta=\cos^2(\al-\be)$ for various $M$ values. 
The other parameters are set to be 
$m_h^{}=120$ GeV, $\tan\beta=2$, and 
$m_H^{}=m_{H^\pm}^{}=m_A^{}=300$ GeV.
}
\label{fig:hhh_tree_loop.eps}
\end{figure}

Here, we study the case in which the condition of $x=0$
(or, $\sin(\al-\be) = -1$) is relaxed. 
When $\sin(\al-\be)$ is much different from $-1$, the renormalized 
couplings are significantly different from their SM values because of 
 the tree level mixing effect\cite{hhh_ext,dec-region}. 
Our main interest is rather the case  
in which the condition $\sin(\al-\be) = -1$ is only slightly relaxed; 
i.e., $\sin(\al-\be) \simeq -1$ or $x \ll 1$. 
We refer such a case as the SM-like regime of the THDM. 
In order to study this case, we introduce the parameter 
$\delta=\cos^2(\al-\be)=1-\sin^2(\al-\be)$ ($\simeq x^2$) 
which directly measures the deviation from the decoupling limit. 
In Figs.~\ref{fig:zzh_tree_loop.eps} and \ref{fig:hhh_tree_loop.eps}, 
we show $(\Delta g_{hZZ}^{THDM}/g_{hZZ}^{})$ and 
$(\Delta \lambda_{hhh}^{THDM}/\lambda_{hhh}^{})$ 
as a function of $\delta$, respectively.
The value of $m_\Phi^{}$ ($=m_H^{}=m_A^{}=m_{H^\pm}^{}$)
is set to be 300 GeV. 
We consider the case of $m_h=120$ GeV and $\tan\beta=2$, and 
the scale $M$ is taken to be $0$, $m_A^{}/2$, $m_A/\sqrt{2}$ and $m_A$.
The solid curves are the results for the one loop corrected 
couplings, and the dotted ones are for the tree level
couplings. 

As shown in Fig.~\ref{fig:zzh_tree_loop.eps}, 
the tree level mixing effect on the $hZZ$ coupling 
is proportional to $\delta$ and the deviation from the SM 
value is negative. 
The non-decoupling effect on the $hZZ$ coupling is insensitive 
to $\delta$ as long as $\delta$ is not large, and 
its deviation is less than one percent to the negative direction. 
The non-decoupling effect becomes maximal for $M=0$, 
and minimum for $M=m_\Phi^{}$. 

Fig.~\ref{fig:hhh_tree_loop.eps} shows that the deviation 
in the tree level $hhh$ coupling can vary due to the Higgs 
mixing effect from $-80$ to $+10$ percent for $\delta = 0.1$, 
depending on the value of $0 < M^2/m_\Phi^2 < 1$.  
The deviation of the tree level $hhh$ coupling vanishes as $\delta=0$,
which reproduces the SM case.
For the fixed value of $\delta$, smaller $M^2$ 
gives larger value (in magnitude) of the tree level $hhh$ coupling.  
At one loop level, the non-decoupling effect 
of the heavy Higgs bosons gives large positive corrections to 
$(\Delta \lambda_{hhh}^{THDM} / \lambda_{hhh}^{SM})$. 
Due to the non-decoupling effect,  
the deviation in the one loop $hhh$ coupling 
can be plus 40 percent for $M=0$, even when  
$\delta=0$. Though the deviation decreases when $\delta$ increases,
such large positive contribution remains for $\delta=0.1$.  
For $M = m_\Phi^{}/2$ the magnitude of the non-decoupling 
effect is smaller than that for $M =0$.  However, 
the deviation can still be larger than 
that induced at tree level by the 
Higgs mixing effect for $0 < M^2/m_\Phi^2 < 1$,  
especially in the region of $0 < \delta < 0.06$.

In conclusion, the large non-decoupling effect of the heavier Higgs bosons 
contributing in loops can be more important than 
the tree level Higgs mixing effect, as long as $\delta$ is not too large.

\subsection{The possible allowed region of the corrections}

\begin{figure}[t]
\begin{center}
\hspace*{-3mm}
\includegraphics[width=8cm,height=6cm]{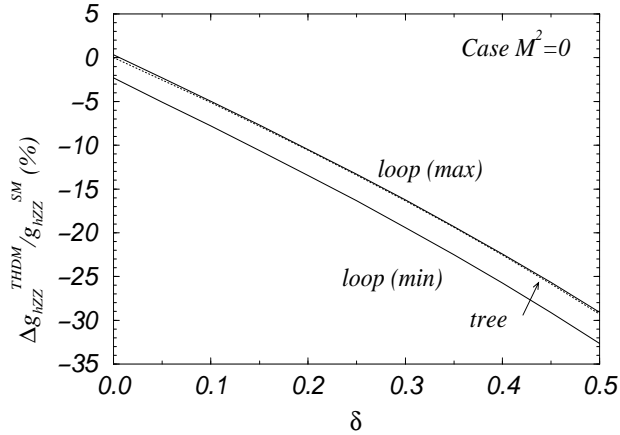}
\end{center}
\vspace*{-8mm}
\caption{
Allowed region of the deviation in the form factor $M_1^{hZZ}$ 
under the constraints obtained from perturbative unitarity and vacuum 
stability as a function of $\delta (=\cos^2(\al-\be))$. 
We set $m_H^{}=m_A^{}=m_{H^\pm}^{}$, and 
the value of $M$ is fixed to be 0 in order to study the maximal
non-decoupling effect.
The upper and lower limits for the one loop renormalized 
couplings are shown as solid curves.  
The value of the tree level one is shown as the dotted curve.
}
\label{fig:m1_scan_M0}
\end{figure}

Finally, we study possible allowed range of the deviation 
in the $hZZ$ and $hhh$ couplings from the SM predictions 
under the experimental and theoretical constraints. 
The free parameters of the Higgs sector in the THDM 
($m_h$, $m_H^{}$, $m_A^{}$, $m_{H^\pm}^{}$, $\alpha$, $\beta$ and $M$) 
are constrained by theoretical consideration as well as 
the available experimental data. 
These are related to the quartic coupling constants in Eq.~(\ref{pot}) 
by Eqs.~(\ref{mass_THDM1}) to (\ref{mass_THDM5}).
We take into account the following bounds in order to 
constrain the parameters. \\
(i) The coupling constants $\lambda_i$ ($i=1-5$) are constrained 
by the requirement of perturbative unitarity\cite{unitarity1}, 
which is described  
by the condition on the S wave amplitudes for the elastic scattering 
of longitudinally polarized gauge bosons as well as 
the Higgs bosons\cite{LQT};
\begin{eqnarray}
  | a^ 0(\varphi_A^{}\varphi_B^{}\to \varphi_C^{}\varphi_D^{}) | < \xi, 
\label{swaveu}
\end{eqnarray}
where $a^0(\varphi_A^{}\varphi_B^{}\to \varphi_C^{}\varphi_D^{})$ 
is the S-wave amplitude for the elastic scattering process 
$\varphi_A^{}\varphi_B^{}\to \varphi_C^{}\varphi_D^{}$ of the 
longitudinally polarized gauge bosons (and Higgs bosons); cf. Appendix D. 
The critical value $\xi$ is a parameter, and 
we here take $\xi=1/2$ in our analysis\cite{HHG}.\\
(ii) The condition of vacuum stability is expressed 
at the tree level by\cite{vacuum-stability} 
\begin{eqnarray}
&&\lambda_1 > 0, \lambda_2 >0, \nonumber\\
&&\sqrt{\lambda_1\lambda_2} + \lambda_3 
+{\rm MIN}(0, \lambda_4+\lambda_5,\lambda_4-\lambda_5) > 0.
\end{eqnarray}
(iii) The LEP precision data imposed strong constraints 
on the radiative corrections to the gauge boson two-point 
functions, which are parameterized by the $S$, $T$ and $U$ 
parameters\cite{peskin-takeuchi}. 
In the THDM, the $T$ parameter 
($\simeq \alpha_{EM}^{-1} \Delta \rho$, where $\Delta \rho (\sim
10^{-3})$ is the deviation of $\rho$ parameter from unity) 
can receive large contributions. 
The analytic formula for $\Delta \rho$ in the THDM is given, 
for example, in Refs.~\cite{rhoTHDM,ST_2HDM}.
To satisfy this constraint, the THDM has to have an
approximate custodial ($SU(2)_V^{}$) symmetry \cite{CSTHDM}.
In the Higgs sector of the THDM, 
there are typically two options for the parameter 
choice in which $SU(2)_V^{}$ is conserved according to the 
assignment of the $SU(2)_V$ charge; 
(1) $m_{H^\pm}^{} \simeq m_A^{}$, and 
(2) $m_{H^\pm}^{} \simeq m_H^{}$ with $\sin^2(\alpha-\beta) \simeq 1$ 
 or $m_{H^\pm}^{} \simeq m_h^{}$ with 
$\cos^2(\alpha-\beta) \simeq 1$ \cite{rhoTHDM,CSTHDM}\footnote{
In terms of the coupling constants, these conditions are 
expressed by (1) $\lambda_4=\lambda_5$, and (2) 
$\lambda_1=\lambda_2=\lambda_3$ with $m_1^2=m_2^2$.}. 
In the present paper, we do not perform a complete scan analysis 
for all the parameter space. 
Instead, we set $m_H^{}=m_A^{}=m_{H^\pm}^{}$ in order to 
reduce the number of parameters. 
By the degeneracy of heavy Higgs bosons, the constraint 
from the $\rho$ parameters is satisfied. 
Then, the free parameters are $m_A^{}$, $\tan\be$, $M$ as well as 
$\delta$ (or $\alpha$). 

\begin{figure}[t]
\begin{center}
\hspace*{-3mm}
\includegraphics[width=8cm,height=6cm]{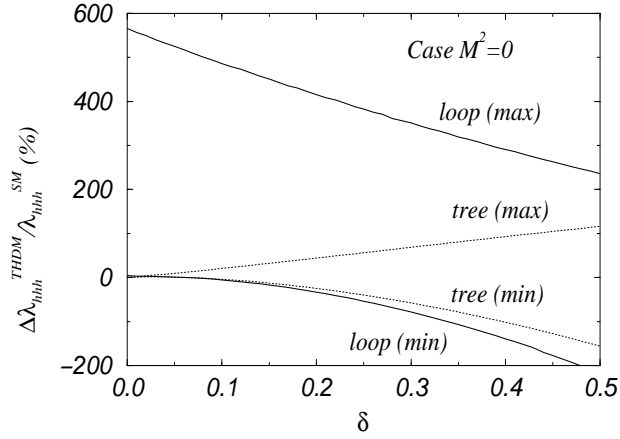}
\end{center}
\vspace*{-8mm}
\caption{
Allowed region of the deviation in the $hhh$ form factor 
under the constraints obtained from perturbative unitarity and vacuum 
stability as a function of $\delta (=\cos^2(\al-\be))$. 
We set $m_H^{}=m_A^{}=m_{H^\pm}^{}$, and 
the value of $M$ is fixed to be 0 in order to study the maximal
non-decoupling effect.
The upper and lower limits for the one loop renormalized 
couplings are shown as solid curves.  
Those for the tree level couplings are shown as dotted curves.
}
\label{fig:hhh_scan_M0}
\end{figure}

\begin{figure}[t]
\begin{center}
\hspace*{-3mm}
\includegraphics[width=8cm,height=6cm]{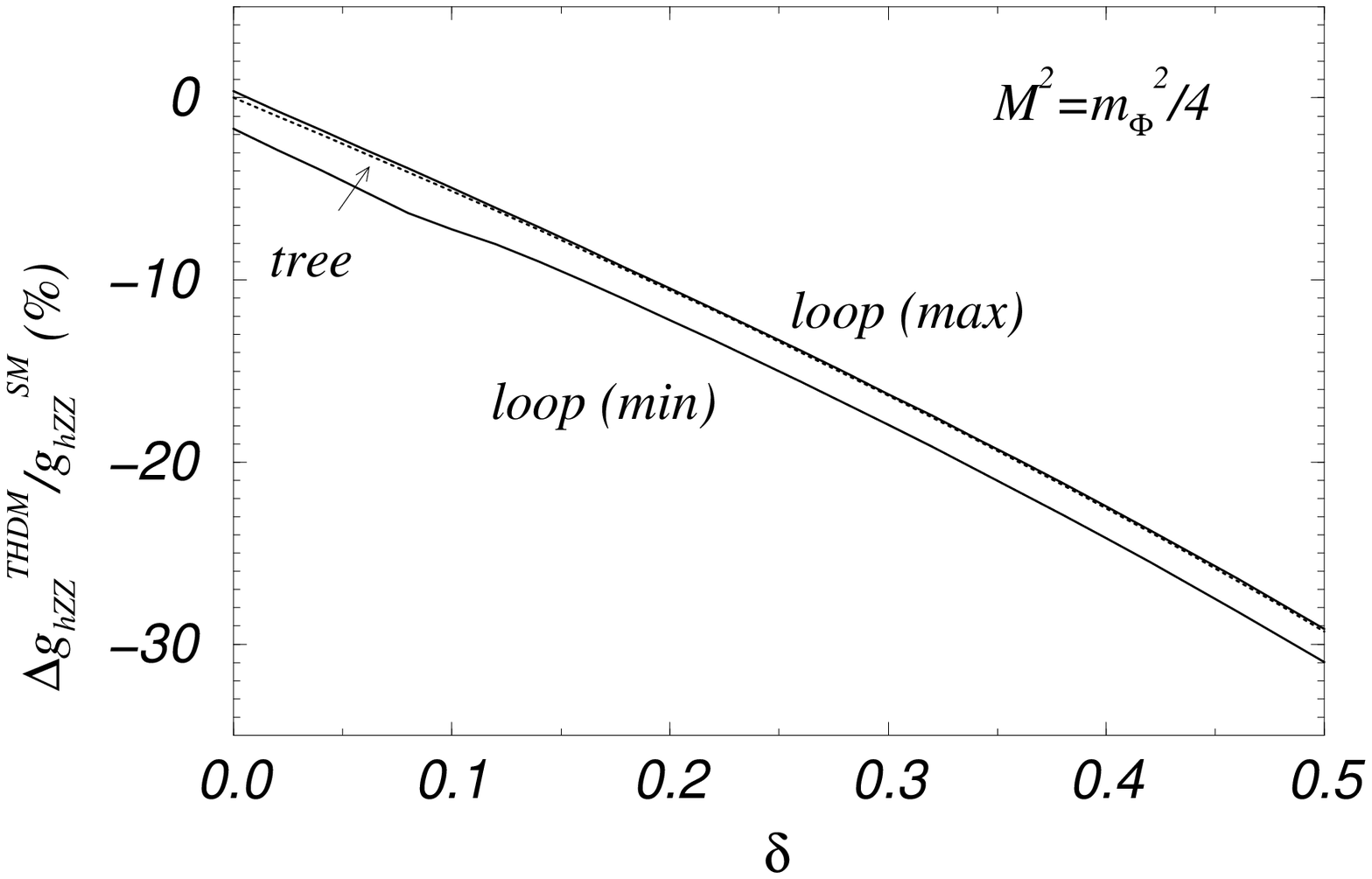}
\end{center}
\vspace*{-8mm}
\caption{
Allowed region of the deviation in the form factor $M_1^{hZZ}$ 
under the constraints obtained from perturbative unitarity and vacuum 
stability as a function of $\delta (=\cos^2(\al-\be))$. 
We set $m_H^{}=m_A^{}=m_{H^\pm}^{} (\equiv m_\Phi^{})$,   
and the value of $M$ is fixed to be $m_\Phi^{}/2$.
The upper and lower limits for the one loop renormalized 
couplings are shown as solid curves.  
The value of the tree level one is shown as the dotted curve.
}
\label{fig:m1_scan_Msq0.5}
\end{figure}
\begin{figure}[t]
\begin{center}
\hspace*{-3mm}
\includegraphics[width=8cm,height=6cm]{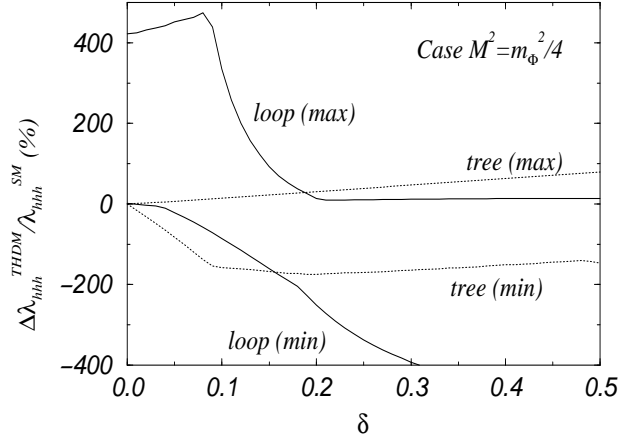}
\end{center}
\vspace*{-8mm}
\caption{
Allowed region of the deviation in the $hhh$ form factor 
under the constraints obtained from perturbative unitarity and vacuum 
stability as a function of $\delta (=\cos^2(\al-\be))$. 
we set $m_H^{}=m_A^{}=m_{H^\pm}^{} (\equiv m_\Phi^{})$, and   
the value of $M$ is fixed to be $m_\Phi^{}/2$.
The upper and lower limits for the one loop renormalized 
couplings are shown as solid curves.  
Those for the tree level couplings are shown as dotted curves.
}
\label{fig:hhh_scan_Msq0.5}
\end{figure}

In Figs.~\ref{fig:m1_scan_M0} and \ref{fig:hhh_scan_M0}, 
we show the allowed region of $(\Delta g_{hZZ}^{THDM}/g_{hZZ}^{SM})$ 
and $(\Delta \lambda_{hhh}^{THDM}/\lambda_{hhh}^{SM})$ for $M=0$
as a function of $\delta$ ($0 < \delta < 0.5$), respectively. 
The mass of the lightest Higgs boson is set to be $m_h=120$ GeV. 
In Fig.~\ref{fig:m1_scan_M0}, we find that the non-decoupling 
loop effect on the $hZZ$ coupling is at most a few percent. 
Due to the leading contribution of the additional Higgs bosons, 
the correction becomes negative. 
Most of the deviation from the SM prediction comes from 
the tree level mixing effect of the factor $\sin^2(\al-\be)$.
On the other hand, as shown in Fig.~\ref{fig:hhh_scan_M0}, 
due to the non-decoupling effect of the heavy Higgs bosons 
the allowed region of $\Delta \lambda_{hhh}^{THDM}/\lambda_{hhh}^{SM}$ 
becomes much wider than that for the tree level. 
The correction can be as large as a few $100$ \%. 
We note that such large deviation from the SM prediction 
cannot be realized solely by the tree level Higgs mixing effect 
for $0 < \delta < 0.5$. 

As the typical case for the non-zero value of $M$, 
we show the result for $M=m_\Phi^{}/2$
in Figs.~\ref{fig:m1_scan_Msq0.5} and \ref{fig:hhh_scan_Msq0.5}, 
where $m_\Phi^{} \equiv m_H^{} = m_A^{} = m_{H^\pm}^{}$. 
All the other parameters are taken to be the same as those in  
Figs.~\ref{fig:m1_scan_M0} and \ref{fig:hhh_scan_M0}. 
As shown, the magnitude of the non-decoupling effect becomes smaller  
as compared to the case with $M=0$,
both in the $hZZ$ and $hhh$ couplings, because of the 
suppression factor $(1-M^2/m_\Phi^{2})^n$; cf. Eqs.~(\ref{m2THDM}) and 
(\ref{m4THDM}). In the $hhh$ coupling, 
the tree level mixing effect becomes significant 
for larger values of $\delta$, 
by which the positive contribution due to the non-decoupling 
effect is canceled: cf. Eq.~(\ref{m4THDM}). 
The deviation from the SM prediction can be larger than a few 
100\% for the small values of $\delta$. 

Some comments are in order related to the unitarity constraint. 
If we take $\xi=1$\cite{LQT} instead of $\xi=1/2$\cite{HHG}, 
the constraint from the perturbative unitarity is relaxed on 
both the tree level and loop effects. 
Then larger values of $m_\Phi^{}$ can be taken,      
so that the possible enhancement due to the 
non-decoupling loop effect becomes greater. 
The rate of enhancement due to the change from $\xi=1/2$ to $\xi=1$ 
is much larger at one loop level than that at tree level. 

\subsection{Discussions}

Before concluding this section, we give a few comments on our analysis.

The large one loop radiative correction of ${\cal O}(1)$ 
to the coupling $\lambda_{hhh}$ in the THDM does not imply the 
breakdown of the perturbative expansion, because the large contribution 
originates from new types of couplings,
e.g., $\lambda_{h\Phi\Phi}^{}$ and $\lambda_{hh\Phi\Phi}$,
that enter in loop calculations. 
Needless to say that we do not expect
such kind of large correction to occur beyond the one loop order. 

We have shown the results by assuming Model II for the Yukawa interaction. 
In the case of Model I, 
our main results presented thus far are essentially unchanged
when $h$ approximately behaves like the SM Higgs boson.   
It is well known that in Model II, the $b \to s \gamma$ 
data imposed a strong constraint on the mass of the 
charged Higgs boson. 
We have not explicitly included the constraint from the 
$b \to s \gamma$ data\cite{bsa-2hdm} in our analysis, 
which can be easily satisfied by assuming that the mass 
of the charged Higgs boson is larger than about 300 GeV.  
In Model I, there is no such strong constraint from 
the $b\to s \gamma$ data. 

The measurement of the $hZZ$ and $hhh$ couplings 
are important not only to confirm the mechanism of the 
electroweak symmetry breaking, but also 
to indirectly explore the property of new physics beyond the SM.
In particular, when the lightest Higgs boson $h$ is 
found to be around 120 GeV at the LHC or LC's, and if its coupling 
with the gauge boson ($hZZ$ or $hWW$) is SM-like, 
the measurement of the $hhh$ 
coupling becomes important to determine the scale 
of the new physics.
If the measured $hhh$ coupling turns out to be much larger than the 
SM prediction and is not possible to be explained by  
the tree level mixing effect in the THDM, we may consider the 
strongly coupled THDM with a relatively low cutoff scale. 
Such large deviation in the $hhh$ coupling 
could be the first indirect signal for the models 
of dynamical symmetry breaking\cite{TC}, or 
models of electroweak baryogenesis\cite{baryogenesis,kos}.
Otherwise, the model with a light $h$ 
should indicate a weakly coupled theory\cite{thdm_RGeffect}.

The trilinear coupling of the lightest Higgs boson
can be measured from studying the scattering processes 
$e^+e^- \to Z^\ast \to Z h^\ast \to Z hh$
and $e^+e^- \to \bar \nu \nu W^{+\ast} W^{-\ast}
            \to \bar \nu \nu  h^\ast
            \to \bar \nu \nu  h  h$ in the $e^+e^-$ collision,
and $\gamma\gamma \to h^\ast \to hh$
at the $\gamma\gamma$ option of the LC.  
In Fig.~4, the momentum dependence on the self-coupling 
has been shown in the SM-like limit.
The corrections turned out to be insensitive for the 
energy below the threshold of the pair 
production of the loop particles\cite{kkosy}. 
Therefore, our main conclusion for the form factors 
of the $hhh$ coupling shown in this paper can be applied 
to the momentum dependent coupling included in the 
above production processes in a good approximation. 
However, the large positive deviation in the $hhh$ coupling 
does not necessarily imply the large deviation in the production 
cross sections by the same rate, because 
the rest of the gauge invariant set of Feynman diagrams usually do not
contain the $hhh$ vertex. 
There have been several studies for the correlation between 
the change of the $hhh$ coupling and 
the production cross sections at the LHC\cite{hhh_LHC} 
and LC's\cite{hhh_yamashita, hhh_ACFA, jikia}. 
From the conclusions in those studies, we expect that 
the large non-decoupling effect in the $hhh$ couplings 
in the THDM can be detected at future experiments at the LC's. 

We finally comment on the case of the MSSM. 
Since the Higgs sector of the MSSM is a special case of the
Model II THDM with $\lambda_i v^2 \simeq {\cal O}(m_W^2)$,
as required by supersymmetry, it belongs to the class of
models in which the heavier Higgs bosons decouple.
Hence, the effect of the Higgs boson loops to
$\Gamma_{hhh}(MSSM)$ is expected to be small.  
A detailed study on this decoupling
behavior of the one loop corrected $hhh$ coupling
in the MSSM can be found in Refs.~\cite{hhh_hollik,lcws02kkosy}. 
We confirmed that our results for large values of $M$ 
are consistent with those in Ref.~\cite{hhh_hollik}.

\section{Conclusion}\label{sec:7}

We have discussed the one loop contributions of the heavy 
additional Higgs bosons 
to the $hZZ$ and $hhh$ couplings in the THDM.
The form factors have been calculated 
in the on-shell scheme, and the deviation from the 
SM predictions are evaluated. 

The renormalized couplings of 
$hZZ$ and $hhh$ can deviate from the SM predictions 
due to two origins: 
the tree level mixing effect between the Higgs bosons 
and the quantum effect of the additional particles 
in the loop.
We found that the deviations in the form factors  
can be large due to the non-decoupling effect of 
the heavy additional Higgs bosons when their masses are predominantly
generated from the vacuum expectation value of the electroweak 
symmetry breaking; i.e., in a strongly-coupled THDM.
In particular, the renormalized $hhh$ coupling 
can largely deviate from the SM prediction 
due to the quartic power term of the masses of the 
heavy Higgs bosons, especially when the mass of the lightest 
Higgs boson $h$ is relatively small. 
Even in the case where approximately only $h$ couples 
to the weak gauge boson so that the mixing 
effect is small, the deviation in the $hhh$ coupling 
from the SM value can be as large as a few 
100\%, 
while that in the $hZZ$ coupling is at most a few times of $-1\%$ 
or less.
When the tree level mixing effect of the Higgs bosons 
is significant, the deviation in the $hZZ$ coupling 
becomes significant by the factor of $\sin(\be-\al)$ 
while the large positive deviation in the $hhh$ coupling 
due to quantum effect is smeared by the mixing effect.
Such large quantum effect on the Higgs tri-linear coupling is 
distinguishable from the Born-level mixing effect, and 
can be detectable at a linear collider. 

In the weakly-coupled THDM, where the masses of the heavy Higgs bosons 
are predominantly generated from the soft breaking mass term $M$, 
the one loop effect is small 
and decouples in the large mass limit.

We have shown how the non-decoupling effect of the 
additional heavy particles in the 
THDM can differ 
the Higgs boson couplings $hZZ$ and $hhh$ 
from the SM prediction. We stress that the quartic mass effect on the effective 
$hhh$ coupling is a general characteristic in any new physics 
model which has the non-decoupling property.

\vspace*{4mm}
\noindent
{\bf ACKNOWLEDGMENTS}

\vspace*{2mm}
\indent
We thank Shingo Kiyoura for useful discussions.
SK and CPY thank the hospitality of National Center for
Theoretical Sciences in Taiwan, ROC, where part of this work was completed.
The work of YO was supported in part by a Grant-in-Aid of the Ministry 
of Education, Culture, Sports, Science, and Technology, Government of
Japan, Nos.~13640309 and 13135225.
The work of CPY was supported in part by the NSF grant PHY-0244919.\\

\appendix

\section{In the Standard Model}\label{app:SM}

Here, we show the calculation for the one loop 
contributions of the top quark 
to the form factors of $hZZ$ and $hhh$ vertices
in the SM. The formulae for the leading top-loop 
contributions in Eqs.~(\ref{sm_hzz_top}) and (\ref{sm_hhh_top}) 
are extracted from the results shown in \ref{app:SM1}. 
The appearance of the quartic mass dependence of the top quark 
in the $hhh$ coupling in Eq.~(\ref{sm_hhh_top})  
can also be shown through the effective potential method 
in \ref{app:SMeff}. 
Here, we consider only the third generation 
quarks (the top and bottom quarks) as the matter fields. 

\subsection{Diagrammatic Method}\label{app:SM1}

The Lagrangian of the SM is given by 
\begin{eqnarray}
   {\cal L}_{SM}&=& + \overline{Q}_L D_\mu \gamma^\mu Q_L  
                    + \overline{t}_R D_\mu \gamma^\mu t_R  
                    + \overline{b}_R D_\mu \gamma^\mu b_R 
\nonumber\\ 
&&                  - \left\{ y_b \overline{Q_L} \Phi b_R 
                    + y_t \tilde{\Phi} Q_L  t_R  + {\rm h.c.} \right\} 
                    + |D^\mu \Phi|^2  - V_{SM},  
\end{eqnarray}
where $Q_L^{}=(t_L^{},b_L^{})^T$, and $D_\mu$ is the covariant
derivative. 
The Higgs potential is defined by 
\begin{eqnarray}
   V_{SM}= -\mu^2 |\Phi|^2 + \lambda |\Phi|^4, \label{pot-sm}
\end{eqnarray}
with the iso-doublet field $\Phi$ being parameterized as 
\begin{eqnarray}
  \Phi = \left[\begin{array}{c}
            w^+ \\
            \frac{1}{\sqrt{2}}(v + h + i z), 
         \end{array}\right], 
\end{eqnarray}
where $v$ ($\simeq 246$ GeV) is the vacuum expectation value,  
$h$ is the Higgs boson, and 
$w^\pm$ and $z$ are the would-be Nambu-Goldstone bosons.

The kinematic term for the Higgs doublet field $\Phi$ 
yields  
\begin{eqnarray}
{\cal L}_{hZZ}^{} =
 \frac{m_Z^2}{v} g_{\mu\nu}^{} Z^\mu Z^\nu h,     
\end{eqnarray} 
where we used the relation 
$m_Z^2 = \frac{g^2+g'^2}{4} v^2$.
The tree level form factors of $hZZ$ vertices are given by 
\begin{eqnarray}
 M_1^{hZZ} &=& \frac{2 m_Z^2}{v}, 
 \hspace*{1cm}M_2^{hZZ} = M_3^{hZZ} = 0. 
\end{eqnarray} 
From the Higgs potential (\ref{pot-sm}), we have 
\begin{eqnarray}
  {V}_{SM} 
    =- T_h  h
    + \frac{1}{2} (m_h^2 - \frac{T_h}{v}) h^2 
    + \frac{m_h^2}{2 v} h^3 + \frac{m_h^2}{8 v^2} h^4 + ... , 
\end{eqnarray}
where we introduced the parameters $T_h$ and $m_h^2$ by 
\begin{eqnarray}
 T_h=  v (\mu^2 - \lambda v^2),  
 \hspace*{1cm}m_h^2 = 2 \lambda v^2, 
\end{eqnarray}  
and eliminated $\mu$ and $\lambda$.
The vacuum condition (the stationary condition) requires 
that the one-point function vanishes at the vacuum. At the 
tree level, this implies that $T_h = 0$, so that the  
parameter $m_h$ denotes the mass of $h$.
The Higgs self-coupling interaction is expressed at tree level by 
\begin{eqnarray}
  \Gamma_{hhh}^{\rm tree} = - \frac{3 m_h^2}{v}, 
\hspace*{1cm}
  \Gamma_{hhhh}^{\rm tree} = - \frac{3 m_h^2}{v^2}.
\end{eqnarray}
The Yukawa interaction generates the mass of the quarks, and 
 $m_t = y_t \frac{v}{\sqrt2}$.
Furthermore, the parameters of the Lagrangian $g, g', v, \lambda, \mu$ and $y_t$ 
can be replaced by 
$m_Z^{}, m_W^{}, v, T_h, m_h^2$ and $m_t$.

The bare parameters of the model can be rewritten in terms of the 
renormalized parameters as 
\begin{eqnarray}
 m_V^2 &\to& m_V^2 + \delta m_V^2,\,\,\,\, (V=W,Z) \\ 
 v &\to& v + \delta v, \\ 
 T_h &\to& T_h + \delta T_h, \\
 m_h^2 &\to& m_h^2 + \delta m_h^2, 
\end{eqnarray}
and the wave function renormalization factors are introduced 
with the renormalized fields by 
\begin{eqnarray}
  Z^\mu &\to& Z_Z^{\frac{1}{2}} Z^\mu + C_{Z\gamma}^{} A^\mu = 
    \left(1 + \frac{1}{2}\delta Z_Z^{} + \cdot\cdot\cdot \right) Z^\mu +
       \left(\delta Z_{Z\gamma}^{} + ... \right) A^\mu, \\
  h &\to& Z_h^{\frac{1}{2}} h = 
    \left(1 + \frac{1}{2}\delta Z_h + \cdot\cdot\cdot \right) h.
\end{eqnarray}
From the kinematic term, we obtain the counter terms for the $hZZ$ 
interaction, 
\begin{eqnarray}
{\cal L}_{hZZ}^{} &\to&
 \frac{m_Z^2}{v} \left( 1 + \frac{\delta m_Z^2}{m_Z^2} - \frac{\delta v}{v} 
 \right)  
g_{\mu\nu}^{} Z^\mu Z^\nu h 
\left(1 + \delta Z_Z^{} + \frac{1}{2} \delta Z_h + \cdot\cdot\cdot \right) 
\nonumber\\     
&\to&
 \frac{m_Z^2}{v}   
g_{\mu\nu}^{} Z^\mu Z^\nu h 
\left(1 + \frac{\delta m_Z^2}{m_Z^2} - \frac{\delta v}{v} +
\delta Z_Z^{} + \frac{1}{2} \delta Z_h \right).
\end{eqnarray} 
We here neglect the effect of the $Z$-$\gamma$ mixing, because 
it is of ${\cal O}(\alpha_{\rm EM}^{})$.
Thus, we have the counter terms for the $hZZ$ form factors as  
\begin{eqnarray}
 \delta M_1^{hZZ} &=& \frac{2 m_Z^2}{v}
\left(\frac{\delta m_Z^2}{m_Z^2} - \frac{\delta v}{v} +
\delta Z_Z^{} + \frac{1}{2} \delta Z_h \right),  
  \hspace*{1cm}\delta M_2^{hZZ} = \delta M_3^{hZZ} = 0, \label{count_hZZ}
\end{eqnarray} 
The bare Higgs Lagrangian can be rewritten as 
\begin{eqnarray}
  {\cal L}_{\rm Higgs} &\to&
     +  \frac{1}{2} p^2    
        \left(1 + \delta Z_h \right) h^2  \nn\\
&&    + \{ (\mu^2-\lambda v^2) \left(1 + \frac{1}{2} \delta Z_h\right) 
       + (\delta \mu^2 - \delta \lambda v^2 - 2 \lambda v \delta v)
      \} \left(v + \delta v \right) h\nn\\
&& - \frac{1}{2} \{ (\mu^3-3\lambda v^2)(1 + \delta Z_h) + 
                    (\delta \mu^2 - 3 \delta \lambda v^2 
                 - 6 \lambda v \delta v) \} h^2\nn\\
&&   - (\lambda v + \delta \lambda v + \lambda \delta v 
                + \frac{3}{2} \lambda v \delta Z_h) h^3
   - \frac{1}{4} (\lambda + \delta \lambda + 2 \lambda \delta Z_h) h^4 \nn\\
&=&
 {\cal L}_{\rm Higgs} 
 + \delta T_h  h 
 + \frac{1}{2} 
   \left\{ (p^2-m_h^2) \delta Z_h - \delta m_h^2 + 
                \frac{\delta T_h}{v} \right\} h^2\nn\\
&& - \left\{ \frac{\delta m_h^2}{2 v} 
           + \frac{m_h^2}{2 v} \left(- \frac{\delta v}{v} 
                                     + \frac{3}{2} \delta Z_h \right) 
   \right\} h^3 \nn\\
&& - \frac{1}{4}\left\{ \frac{\delta m_h^2}{2 v^2} 
                 +  \frac{m_h^2}{v^2}\left(- \frac{\delta v}{v} 
                                          + \delta Z_h \right) \right\} h^4, 
\end{eqnarray}
where $\delta T_h$ and $\delta m_h^2$ are 
consistently related to the shift of the Lagrangian parameters 
$\mu$ and $\lambda$ by 
\begin{eqnarray}
\delta T_h &\equiv& 
(\delta \mu^2 - \delta \lambda v^2 - 2 \lambda v \delta v) v
            = \delta \left\{v (\mu^2 - \lambda v^2) \right\}, \\
\delta m_h^2 &\equiv& 2 \delta \lambda v^2 + 4 \lambda v \delta v     
          = \delta \left( 2 \lambda v^2 \right). 
\end{eqnarray}
The counter terms for the $hhh$ and $hhhh$ vertices are 
\begin{eqnarray}
\delta \Gamma_{hhh} &=& - \left\{ \frac{3\delta m_h^2}{v} 
           + \frac{3 m_h^2}{v} \left(- \frac{\delta v}{v} 
                                     + \frac{3}{2} \delta Z_h \right) 
   \right\}, \label{count_hhh}\\ 
\delta \Gamma_{hhhh} &=& - \left\{ \frac{3 \delta m_h^2}{v^2} 
                 +  \frac{6 m_h^2}{v^2}\left(- \frac{\delta v}{v} 
                                          + \delta Z_h \right) \right\}. 
\label{count_hhhh}
\end{eqnarray}
Based on Eqs.~(\ref{count_hZZ}) and (\ref{count_hhh}),  
we have to provide the counter term parameters 
$\delta m_h^2$, $\delta Z_h^{}$, $\delta v$, 
$\delta m_Z^2$, $\delta Z_Z^{}$ 
in order to calculate the one loop form factors of $hZZ$ and $hhh$.
In the following, we determine all these parameters by imposing proper  
renormalization conditions. 

The renormalization is performed in the on-shell scheme\cite{on-shell}.
The counter terms of the gauge boson masses ($\delta m_W^2$, $\delta m_Z^2$) 
and wave functions ($\delta Z_W^{}$, $\delta Z_Z^{}$)
are obtained by calculating the transverse part $\Pi_T^{VV}(p^2)$ 
of the two-point function 
\begin{eqnarray}
 \Pi_{\mu\nu}^{VV}(p^2) = 
 \left(- g_{\mu\nu} + \frac{p_\mu p_\nu}{p^2}  \right) \Pi_T^{VV}(p^2) 
  + \frac{p_\mu p_\nu}{p^2}  \Pi_L^{VV}(p^2), 
\end{eqnarray}
where $VV=WW$ or $ZZ$.
In the one-shell renormalization scheme, we obtain 
\begin{eqnarray}
  \delta m_Z^2 &=& {\rm Re} \Pi^{ZZ({\rm 1PI})}_T(m_Z^2), \\ 
  \delta Z_Z   &=& - \left.\frac{\partial}{\partial p^2 }
                     {\rm Re} \Pi^{ZZ({\rm 1PI})}_T(p^2) 
              \right|_{p^2=m_Z^2}. 
\end{eqnarray}
We define $\delta v$ by
\begin{eqnarray}
\frac{\delta v}{v} = \frac{1}{2}\,  
\frac{1}{m_W^2} {\rm Re} \Pi^{WW({\rm 1PI})}_T(0) + ({\rm vertex}\,\,
{\rm and}\,\, {\rm box}\,\,
  {\rm corrections}) \, , \label{dv}
\end{eqnarray}
where the ``(vertex and box corrections)'' 
in Eq.~(\ref{dv}) is ${\cal O}(\alpha)$, 
and is neglected in our calculations.

The other counter terms 
$\delta T_h$, $\delta m_h^2$ and $\delta Z_h$ are determined in the 
following way.
Firstly, the tadpole must be zero after renormalization; i.e., 
at tree level we demand $T_h=v (\mu^2 - \lambda v^2)=0$, and at one loop 
level, we impose  
\begin{eqnarray}
  \Gamma_{h}^{R} \equiv  T_h^{\rm tree} + T_{h}^{\rm 1PI} + \delta T_h = 0,  
\end{eqnarray}
where $\Gamma_{h}^{R}$ is the renormalized tadpole and 
$T_{h}^{1PI}$ is the one loop Feynman diagram of the tadpole 
(cf. Appendix~\ref{app:1PI_SM}). 
This determines the counter-term $\delta T_h$ as 
\begin{eqnarray}
  \delta T_h = - T_{h}^{1PI}.   
\end{eqnarray}
Secondly, we determine the rest of the counter terms in the on-mass shell 
scheme, i.e., 
\begin{eqnarray}
  {\rm Re} \Gamma_{hh}^{R}[m_h^2] &=& 0,\\
  \left.\frac{\partial}{\partial p^2} 
  {\rm Re} \Gamma_{hh}^{R}[p^2]\right|_{p^2=m_h^2} &=& 1,
\end{eqnarray}
where $\Gamma_{hh}^{R}[p^2]$ is the renormalized 
two-point function of $hh$:
\begin{eqnarray}
  \Gamma_{hh}^{R}[p^2]
  = (p^2-m_h^2) (1 + \delta Z_h) 
    - \delta m_h^2 + \frac{\delta T_h}{v} 
    +\Pi_{hh}^{1PI}(p^2),       
\end{eqnarray}
and  
  $\Pi_{hh}^{1PI}(p^2)$
is the 1PI Feynman diagram contribution  (cf. Appendix~\ref{app:1PI_SM}).
Therefore, we obtain 
\begin{eqnarray}
 \delta m_h^2&=& + {\rm Re} \Pi_{hh}^{1PI}(m_h^2) 
                 - \frac{1}{v} {\rm Re}  T_{h}^{1PI} ,\\   
 \delta Z_h  &=&-\left.\frac{\partial}{\partial p^2} 
             {\rm Re}  \Pi_{hh}^{1PI}(p^2)\right|_{p^2=m_h^2}.      
\end{eqnarray}

Using the counter terms 
$\delta m_Z^{}$, $\delta Z_Z^{}$,  
$\delta v$, $\delta T_h$, $\delta m_h^2$, and $\delta Z_h$ which are 
determined above with the 1PI diagrams listed in 
Appendix~\ref{app:1PI_SM}, 
we obtain the renormalized 
form factors of $hZZ$ and $hhh$. 
Consequently, the renormalized form factors are given by 
\begin{eqnarray}
 M_i^{hZZ}(p_1^2,p_2^2,q^2) &=&  M_i^{hZZ(\rm tree)} + 
                M_i^{hZZ(\rm 1PI)} + \delta M_i^{hZZ},  (i=1-3) 
\label{sm_hzz}\\
 \Gamma_{hhh}^{}(p_1^2,p_2^2,q^2) &=& \Gamma_{hhh}^{\rm tree} 
                +\Gamma_{hhh}^{\rm 1PI} + \delta \Gamma_{hhh}^{},  
\label{sm_hhh}
\end{eqnarray}
where the momentum $q^\mu$ in Eq.~(\ref{sm_hzz}) is that 
of the external Higgs boson line.
The leading contributions of the top quark mass in 
Eqs.~(\ref{sm_hzz_top}) and (\ref{sm_hhh_top}) 
can be obtained from Eqs.~(\ref{sm_hzz}) and (\ref{sm_hhh}) 
by taking the large $m_t$ limit in the explicit expressions 
of each diagram contribution listed in \ref{app:1PI_SM} 

\subsection{Effective potential method}\label{app:SMeff}

The quartic power dependence of the top quark mass can be 
reproduced in the effective potential method. 
The effective potential provides the information of vertex 
functions with zero external momenta at each loop level. 
The one loop effective potential is given by 
\begin{eqnarray}
   V_{eff}[\varphi]= V_{\rm tree}[\varphi] 
                   + \frac{1}{64\pi^2} 
                     N_{c_f} N_{s_f} (-1)^{2 s_f}  
                     M_f^4[\varphi]
             \left(\ln\left[\frac{M_f^2}{Q^2}\right]-\frac{3}{2}\right),  
\label{eff}
\end{eqnarray}
where $\varphi=\langle \phi \rangle = v + \langle h \rangle$, 
$N_{c_f}$ is the color number,   
$s_f$ ($N_{s_f}$) is the spin (degree of freedom) of the field $f$  
in the loop, $M_f[\varphi]$ is the field dependent mass of $f$, 
and $Q$ is an arbitrary scale. 

The bare parameters $\mu^2$ and $\lambda$ in the SM 
Higgs potential $V_{\rm SM}^{}[\varphi]$ in Eq.~(\ref{pot-sm})
($=V_{\rm tree}[\varphi]$)  
can be eliminated after introducing the one loop 
corrected vacuum expectation value $v$ and the mass $m_h$ 
in the following conditions:
\begin{eqnarray}
   \left.
   \frac{\partial}{\partial \varphi} V_{eff}[\varphi] 
      \right|_{\varphi=v} &=& 0, \label{cond1}\\
 \left.
   \frac{\partial^2}{\partial \varphi^2} V_{eff}[\varphi] 
\right|_{\varphi=v} &=& m_h^2. \label{cond2} 
\end{eqnarray}
Let us consider the top-quark loop effect.
Namely, in Eq.~(\ref{eff}), $f=t$, $N_{C_t}=3$, $N_{S_t}=2$ and the 
field dependent mass of $t$ is given by 
\begin{eqnarray}
    M_t[\varphi] = y_t \frac{\varphi}{\sqrt{2}}.  
\end{eqnarray}
The result  (\ref{sm_hhh_top})
of the renormalized coupling constant $\Gamma_{hhh}^{SM}$ 
is then obtained from 
\begin{eqnarray}
   \left.
   \frac{\partial^3}{\partial \varphi^3} V_{eff}[\varphi] 
      \right|_{\varphi=v} = 3 \frac{m_h^2}{v}
       \left\{ 1 - \frac{N_{c_t}}{3 \pi^2} \frac{m_t^4}{v^2 m_h^2}
         \right\}, \label{smhhh}
\label{mt4term_hhh_SM}
\end{eqnarray}
where $m_t$ is the mass of the top quark. 
Because the top quark is a fermion, its loop effect on the effective 
potential is negative.  

Let us examine the origin of this quartic mass contribution. 
If we write $\varphi=v_0 + h$, then the effective potential is 
\begin{eqnarray}
  V_{eff}&=& -\frac{\mu^2}{2} (v_0+h)^2 
                    + \frac{1}{4} \lambda (v_0+h)^4 
      - \frac{N_c}{16\pi^2} \frac{y_t^4}{4} (v_0+h)^4 
    \left[ \ln \frac{y_t^2 v_0^2 \left(1+\frac{h}{v_0}\right)^2}{2 Q^2}
        - \frac{3}{2}\right]\nonumber\\ 
&=& -\frac{\mu^2}{2} (v_0+h)^2 
                    + \frac{1}{4} \tilde{\lambda}  (v_0+h)^4 
- \frac{N_c}{16\pi^2} \frac{y_t^4}{2} v_0^4 
   \left(\frac{h}{v_0} 
  + \frac{7}{2}\frac{h^2}{v_0^2}+ \frac{13}{3}\frac{h^3}{v_0^3} 
  + \cdot\cdot\cdot \right), 
\end{eqnarray}
where 
\begin{eqnarray}
\tilde{\lambda} &\equiv&   \lambda 
      - \frac{N_c}{16\pi^2} y_t^4 
  \left( \ln \frac{y_t^2 v_0^2}{2 Q^2}-\frac{3}{2} \right).
\end{eqnarray}
The first to third derivatives are calculated as
\begin{eqnarray}
\frac{\partial V_{eff}}{\partial h} &=& 
 - \mu^2 v_0 + \tilde{\lambda} v_0^3 
 - \frac{1}{2} \frac{N_c}{16\pi^2} y_t^4 v_0^3,\\
\frac{\partial^2 V_{eff}}{\partial h^2} &=& 
 - \mu^2 + 3 \tilde{\lambda} v_0^2
 - \frac{7}{2} \frac{N_c}{16\pi^2} y_t^4 v_0^2,\\
\frac{\partial^3 V_{eff}}{\partial h^3} &=& 
 6 \tilde{\lambda} v_0 
 - 13 \frac{N_c}{16\pi^2} y_t^4 v_0.
\end{eqnarray}
Using Eqs.~(\ref{cond1}) and (\ref{cond2}), 
we can eliminate $\mu^2$ and $\tilde{\lambda}$ by introducing 
the renormalized (at zero momentum) mass $m_h^2$.  
Then, we obtain the renormalized coupling  
$\lambda_{hhh}^{SM}$, as given in Eq.~(\ref{smhhh}).   
The logarithmic term in the effective potential 
is completely eliminated by the mass renormalization. 
On the contrary, the higher dimensional operator terms 
in the effective potential generally survive even after 
the mass renormalization, which yield the ${\cal O}(m_t^4)$ 
correction to the tree level $hhh$ coupling.

\subsection{1PI diagram contributions in the SM}\label{app:1PI_SM}

We list the relevant one-particle irreducible (1PI) $n$-point functions
for $n=1,2,3$. 
The calculation is performed in Landau gauge, so that the 
Nambu-Goldstone bosons are  massless ($m_{w^\pm}^{}=m_z^{}=0$).
The SM Higgs boson coupling constants to be used below are defined as
by
\begin{eqnarray}
\nonumber\lambda_{hhh}^{ }v =
\nonumber\lambda_{hzz}^{ }v =
\nonumber\frac{1}{2}  \lambda_{hw^+w^-}^{ }v =
\nonumber 4 \lambda_{hhhh}^{ }v^2 =
\nonumber 2 \lambda_{hhzz}^{ }v^2 =
\nonumber \lambda_{hhw^+w^-}^{ }v^2 = -\frac{m_h^2}{2}.
\end{eqnarray}
 
\subsubsection{One- and two-point functions}

The one loop top-bottom contributions and the Higgs scalar contributions 
to $\Pi_{T}^{ZZ}(p^2)$ and $\Pi_{T}^{WW}(p^2)$ are given by 
\begin{eqnarray}
\Pi_{T}^{ZZ}(p^2) &=& 
- \frac{N_c}{16\pi^2} \frac{16 m_Z^2}{v^2}
\left[ (\frac{1}{2} I_f^2-I_fQ_f s_W^2 + Q_f^2s_W^4) 
\left\{ (D-2)B_{22} +p^2 (B_1+B_{21})\right\} \right.\nonumber\\&&
\left.+\frac{}{}(I_f-Q_fs_W^2)Q_fs_W^2m_f^2B_0\right](p^2,m_f,m_f)
\nonumber\\  &&
- \frac{N_c}{16\pi^2} \frac{m_Z^2}{v^2} 
 \left\{ c_{2W}^{} B_5(p^2,m_w,m_w) + B_5(p^2,m_z,m_h) \right\}
\\  
\Pi_{T}^{WW}(p^2) &=& 
- \frac{N_c}{16\pi^2} \frac{m_W^2}{v^2}
  \left\{  (D-2) 4 B_{22} + 4 p^2 (B_1 + B_{21})\right\}(p^2;m_t,m_b)   
\nonumber\\  &&
- \frac{N_c}{16\pi^2} \frac{m_W^2}{v^2} 
 \left\{ B_5(p^2;m_w,m_z) + B_5(p^2;m_w,m_h) \right\}, 
\end{eqnarray}
where we used the Passarino-Veltman functions\cite{pv} for the tensor
coefficients of the loop integrals, and we define
$B_5(p^2;m_1^{},m_2^{}) = 
   A(m_1^{})+A(m_2^{}) - 4 B_{22}(p^2;m_1^{},m_2^{})$.
$I_f$ and $Q_f$ are the isospin quantum number and the electric charge
of the fermion $f$. For example, $I_f=1/2$ and $Q_f=2/3$ for $f=t$.
Also, $s_W=\sin \theta_W$, where $\theta_W$ is the weak-mixing angle. 
The 1PI tadpole contributions are calculated as 
\begin{eqnarray}
 T_h^{1PI} &=&  \sum_{f=t,b} \left\{
- \frac{N_c}{16\pi^2} \frac{4 N_c m_f^2}{v} A(m_f)
\right\}  
-\frac{1}{16\pi^2} 3 \lambda_{hhh}^{ } A(m_h).
\end{eqnarray}
The 1PI diagram contributions to the Higgs boson two-point function 
is obtained as 
\begin{eqnarray}
 \Pi_{hh}^{1PI}(p^2) &=&  \sum_{f=t,b} \left[
- \frac{N_c}{16\pi^2} \frac{N_c m_f^2}{v^2} 
\left\{ 4 A(m_f) + (-2p^2+8 m_f^2) B_0(p^2;m_f,m_f) 
\right\}\right] \nonumber\\&&
+ \frac{1}{16\pi^2} \left\{
+(\lambda_{h\omega\omega}^{ })^2  
B_0(p^2; m_\omega,m_\omega)
+2(\lambda_{hzz}^{ })^2 
B_0(p^2; m_z,m_z) \frac{}{} \right.
\nonumber\\&&
\left.\frac{}{}+18 (\lambda_{hhh}^{ })^2 
B_0(p^2; m_h,m_h)
- 12 \lambda_{hhhh}^{ }
A(m_h)
 \right\}.
\end{eqnarray}

\subsubsection{The $hZZ$ form factors} 

The 1PI diagrams of the top-loop contribution 
to $M_1^{hZZ}(p_1^2,p_2^2,q^2)$ is calculated as  
\begin{eqnarray}
&&M_1^{hZZ}(p_1^2,p_2^2,q^2)= \nonumber\\
&&
+\frac{1}{16\pi^2} \frac{32 N_c m_f^2 m_Z^2}{v^3} 
\left[
\left\{
\frac{1}{2}I_f^2 - I_f Q_f s_W^2 +Q_f^2 s_W^4
\right\} 
\left\{ \frac{}{}
2 p_1^2 C_{21}+2 p_2^2 C_{22} +4 p_1p_2 C_{23} 
\right.\right. \nonumber\\&&
\left.\left.
+ 2 (D-2) C_{24}
+(3p_1^2+p_1p_2) C_{11} 
+ (3p_1p_2 + p_2^2) C_{12} + (p_1^2+p_1p_2) C_0 \frac{}{}
\right\} 
\right. \nonumber \\
&&
\left.
+ \left( I_fQ_fs_W^2-Q_f^2s_W^4 \right) 
\left\{ \frac{}{}
p_1^2 C_{21} + p_2^2 C_{22} +2 p_1p_2 C_{23} + D C_{24} 
\right.\right. \nonumber\\
&& \left.\left.
+ (p_1^2+p_1p_2) C_{11} + (p_1p_2+p_2^2) C_{12} +m_f^2 C_0
\frac{}{}\right\}
\right] (p_1^2,p_2^2,q^2; m_f, m_f, m_f)\nonumber\\
&&
+\frac{1}{16\pi^2} \frac{m_Z^2}{v^2} 
\left[ \frac{}{}
2 \cos^22\theta_W  \lambda_{hw^+w^-}^{} 
B_0 (q; m_{w^\pm}^{},m_{w^\pm}^{}) 
+
2 \lambda_{hzz}^{} 
B_0 (q; m_{z}^{},m_{z}^{}) 
\right. \nonumber\\&& 
+6 \lambda_{hhh}^{} 
B_0 (q; m_{h}^{},m_{h}^{}) 
-8 \cos^22\theta_W \lambda_{hw^+w^-}^{} 
C_{24} (p_1^2,p_2^2,q^2; m_{w^\pm}^{},m_{w^\pm}^{},m_{w^\pm}^{}) 
\nonumber\\&& \left. 
-8 \lambda_{hzz}^{} 
C_{24} (p_1^2,p_2^2,q^2; m_{z}^{},m_{h}^{},m_{z}^{}) 
-24 \lambda_{hhh}^{} 
C_{24} (p_1^2,p_2^2,q^2; m_{h}^{},m_{z}^{},m_{h}^{}) 
\frac{}{}\right], 
\end{eqnarray}
\begin{eqnarray}
&&M_2^{hZZ}(p_1^2,p_2^2,q^2)= 
\nonumber\\&&
-\frac{1}{16\pi^2} \frac{32 N_c m_f^2 m_Z^4}{v^3} 
\left[
\left\{
\frac{1}{2}I_f^2 - I_f Q_f s_W^2 +Q_f^2 s_W^4
\right\} 
\left\{ \frac{}{}
4 C_{23}+3 C_{12} + C_{11} + C_0 \right\} 
\right. \nonumber \\
&&
\left.
+ \left( I_fQ_fs_W^2-Q_f^2s_W^4 \right) 
\left\{ \frac{}{}
 C_{11} -  C_{12} \right\}
\right] (p_1^2,p_2^2,q^2; m_f, m_f, m_f)\nonumber\\
&&
+\frac{1}{16\pi^2} \frac{m_Z^4}{v^2} 
\left[ \frac{}{}
-8 \cos^22\theta_W \lambda_{hw^+w^-}^{} 
C_{1223} (p_1^2,p_2^2,q^2; m_{w^\pm}^{},m_{w^\pm}^{},m_{w^\pm}^{}) 
\right. \nonumber\\&& \left.
- 8 \lambda_{hzz}^{} 
C_{1223} (p_1^2,p_2^2,q^2; m_{z}^{},m_{h}^{},m_{z}^{}) 
-24 \lambda_{hhh}^{} 
C_{1223} (p_1^2,p_2^2,q^2; m_{h}^{},m_{z}^{},m_{h}^{}) 
\frac{}{}\right], 
\end{eqnarray}
where $C_{1223}=C_{12}+C_{23}$, and 
\begin{eqnarray}
&& \!\!\!\!M_3^{hZZ}(p_1^2,p_2^2,q^2)= 
\nonumber\\&&
\!\!\!\!-\frac{1}{16\pi^2} \frac{32 N_c m_f^2 m_Z^4}{v^3} 
\left( \frac{1}{2}I_f^2 - I_f Q_fs_W^2\right) 
\left\{ \frac{}{}
 C_{12} -  C_{11} - C_0 \right\}
(p_1^2,p_2^2,q^2; m_f, m_f, m_f).
\end{eqnarray}

\subsubsection{The 1PI $hhh$ form factor}

The 1PI top-loop contribution to the $hhh$ coupling is 
calculated as  
\begin{eqnarray}
&&\Gamma_{hhh}^{1PI}(p_1^2,p_2^2,q^2)=\nonumber\\
&&-\!\!\sum_{f=t,b} \left[
\frac{1}{16\pi^2} \frac{8 N_c m_f^4}{v^3} 
\left\{ \frac{}{}
3 \left(
p_1^2 C_{21} + p_2^2 C_{22} 
+2 p_1p_2 C_{23} + D C_{24}
\right) 
+\left( 4 p_1^2 + 2 p_1p_2 \right) C_{11}
\right. \right. 
\nonumber\\
&& \left.\left.
+\left( 2 p_2^2 + 4 p_1p_2 \right) C_{12} 
+ \left(m_f^2 + p_1^2+p_1p_2\right) C_0
\right\} \right]
(p_1^2,p_2^2,q^2; m_f, m_{f}, m_f)
\nonumber\\&& 
+ \frac{1}{16\pi^2} 
\left[ 
+2 \lambda_{hw^+w^-}^{}\lambda_{hhw^+w^-}^{} 
\left\{\frac{}{}
B_0 (q^2; m_{w^\pm}^{},m_{w^\pm}^{}) 
+B_0 (p_1^2; m_{w^\pm}^{},m_{w^\pm}^{})
+B_0 (p_2^2; m_{w^\pm}^{},m_{w^\pm}^{}) 
\frac{}{}\right\} 
\right.\nonumber \\&& \left.
-2 \lambda_{hw^+w^-}^{3} 
C_0 (p_1^2,p_2^2,q^2; m_{w^\pm}^{},m_{w^\pm}^{},m_{w^\pm}^{})
\right.\nonumber \\&& \left.
+4 \lambda_{hzz}^{}\lambda_{hhzz}^{} 
\left\{\frac{}{}
 B_0 (q^2; m_{z}^{},m_{z}^{}) 
+B_0 (p_1^2; m_{z}^{},m_{z}^{}) 
+B_0 (p_2^2; m_{z}^{},m_{z}^{}) 
\frac{}{}\right\} 
\right.\nonumber \\&& \left.
-4 \lambda_{hzz}^{3} 
C_0 (p_1^2,p_2^2,q^2; m_{z}^{},m_{z}^{},m_{z}^{})
\right.\nonumber \\&& \left.
+72 \lambda_{hhh}^{}\lambda_{hhhh}^{} 
\left\{\frac{}{}
 B_0 (q^2; m_{h}^{},m_{h}^{}) 
+B_0 (p_1^2; m_{h}^{},m_{h}^{}) 
+B_0 (p_2^2; m_{h}^{},m_{h}^{}) 
\frac{}{}\right\} 
\right.\nonumber \\&& \left.
-108 \lambda_{hhh}^{3} 
C_0 (p_1^2,p_2^2,q^2; m_{h}^{},m_{h}^{},m_{h}^{})
\right].
\end{eqnarray}

\section{In the Two Higgs Doublet Model}\label{app:1PI_THDM}

The kinetic and mass terms of the bare Higgs Lagrangian is written 
in terms of the renormalized quantities and the counter-term parameters as 
\begin{eqnarray}
 {\cal L}_{\rm Higgs} 
&\to&
 {\cal L}_{\rm Higgs} \nonumber\\ 
&& + \delta T_h^{}  h  + \delta T_H^{}  H  \nonumber\\
&& + \frac{1}{2} 
   \left\{ (p^2-m_h^2) \delta Z_h - \delta m_h^2  
    +\frac{\sin^2\al}{\cos\be} \frac{\delta T_1}{v} 
    +\frac{\cos^2\al}{\sin\be} \frac{\delta T_2}{v} 
\right\} h^2\nonumber\\
&& + \frac{1}{2} 
   \left\{ (p^2-m_H^2) \delta Z_H^{} - \delta m_H^2  
    +\frac{\cos^2\al}{\cos\be} \frac{\delta T_1}{v} 
    +\frac{\sin^2\al}{\sin\be} \frac{\delta T_2}{v} 
\right\} H^2\nonumber\\
&& + 
   \left\{ \frac{}{}(2 p^2-m_h^2 - m_H^2) \delta C_h 
          - (m_H^2 - m_h^2) \delta \alpha  
\right.
\nonumber \\ &&\left. \hspace*{2cm}
    + \cos\al\sin\al \left(
    - \frac{1}{\cos\be} \frac{\delta T_1}{v} 
    +\frac{1}{\sin\be} \frac{\delta T_2}{v} 
 \right)
\right\} H h \nonumber\\
&& + \frac{1}{2} 
   \left\{ p^2 \delta Z_z 
    + {\cos\be} \frac{\delta T_1}{v} 
    + {\sin\be} \frac{\delta T_2}{v} 
\right\} z^2\nonumber\\
&& + \frac{1}{2} 
   \left\{ (p^2-m_A^2) \delta Z_A^{} - \delta m_A^2  
\right.
\nonumber\\
&&
\left.   \hspace*{1cm}
 +\left(\frac{\sin^2\be}{\cos\be}-\cos\be +\frac{1}{\cos\be} \right) 
                                                  \frac{\delta T_1}{2v} 
    +\left(\frac{\cos^2\be}{\sin\be}-\sin\be +\frac{1}{\sin\be} \right) 
                                                  \frac{\delta T_2}{2v} 
\right\} A^2\nonumber\\
&& + 
   \left\{ \frac{}{}(2 p^2-m_A^2) \delta C_A  + m_A^2 \delta \beta  
    - \sin\be \frac{\delta T_1}{v} 
    + \cos\be \frac{\delta T_2}{v} 
\right\} z A \nonumber\\
&& + 
   \left\{ p^2 \delta Z_{w^\pm}^{} 
    + {\cos\be} \frac{\delta T_1}{v} 
    + {\sin\be} \frac{\delta T_2}{v} 
\right\} w^+w^-\nonumber\\
&& + 
   \left\{ (p^2-m_{H^\pm}^2) \delta Z_{H^\pm}^{} - \delta m_{H^\pm}^2  
\right.
\nonumber\\
&&
\left.   \hspace*{1cm}
 +\left(\frac{\sin^2\be}{\cos\be}-\cos\be +\frac{1}{\cos\be} \right) 
                                                  \frac{\delta T_1}{2v} 
    +\left(\frac{\cos^2\be}{\sin\be}-\sin\be +\frac{1}{\sin\be} \right) 
                                                  \frac{\delta T_2}{2v} 
\right\} H^+H^-
\nonumber\\
&& + 
   \left\{ \frac{}{}(2 p^2-m_{H^\pm}^2) \delta C_{H^+}  + m_{H^\pm}^2 \delta \beta  
    - \sin\be \frac{\delta T_1}{v} 
    + \cos\be \frac{\delta T_2}{v} 
\right\} (w^+H^-+H^+w^-),  
\end{eqnarray}
where 
\begin{eqnarray}
     \delta T_1 &=& \cos\al \delta T_H^{} - \sin\al \delta T_h^{}, \\
     \delta T_2 &=& \sin\al \delta T_H^{} + \cos\al \delta T_h^{}.  
\end{eqnarray} 

\subsection{One- and two-point functions}

The explicit expressions for 
the relevant 1PI diagrams are given in terms 
of the Passarino-Veltman functions\cite{pv} below. 
The Yukawa couplings are assumed to be of the Model II THDM\cite{HHG}.
\begin{eqnarray}
\Pi_{T}^{ZZ}(p^2) &=& 
- \frac{N_c}{16\pi^2} \frac{16 m_Z^2}{v^2}
\left[ (\frac{1}{2} I_f^2-I_fQ_f s_W^2 + Q_f^2s_W^4) 
\left\{ (D-2)B_{22} +p^2 (B_1+B_{21})\right\} \right.\nonumber\\&&
\left.+\frac{}{}(I_f-Q_fs_W^2)Q_fs_W^2m_f^2B_0\right](p^2,m_f,m_f)
\nonumber\\  &&
- \frac{N_c}{16\pi^2} \frac{m_Z^2}{v^2} 
 \left\{ c_{2W}^{} B_5(p^2,m_w,m_w) + B_5(p^2,m_z,m_h) \right\},
\\  
\Pi_{T}^{WW}(p^2) &=& 
- \frac{N_c}{16\pi^2} \frac{m_W^2}{v^2}
  \left\{  (D-2) 4 B_{22} + 4 p^2 (B_1 + B_{21})\right\}(p^2;m_t,m_b)   
\nonumber\\  &&
- \frac{N_c}{16\pi^2} \frac{m_W^2}{v^2} 
 \left\{ B_5(p^2;m_w,m_z) + B_5(p^2;m_w,m_h) \right\}.
\end{eqnarray}

\begin{eqnarray}
 T_h^{1PI} &=&   
- \frac{N_c}{16\pi^2} \frac{4 m_t^2}{v} \frac{\cos\al}{\sin\be}A(m_t)
+ \frac{N_c}{16\pi^2} \frac{4 m_b^2}{v} \frac{\sin\al}{\cos\be}A(m_b)
\nonumber\\&&
-\frac{1}{16\pi^2} 
\left\{
\lambda_{hH^+H^-}^{} A(m_{H^\pm}^{})
+\lambda_{hAA}^{} A(m_A^{})
+3 \lambda_{hhh}^{} A(m_{h}^{})
+ \lambda_{hHH}^{} A(m_{H}^{})
\right\}, \\
 T_H^{1PI} &=&   
- \frac{N_c}{16\pi^2} \frac{4 m_t^2}{v} \frac{\sin\al}{\sin\be}A(m_t)
- \frac{N_c}{16\pi^2} \frac{4 m_b^2}{v} \frac{\cos\al}{\cos\be} A(m_b)
\nonumber\\&&
-\frac{1}{16\pi^2} 
\left\{
\lambda_{HH^+H^-}^{} A(m_{H^\pm}^{})
+\lambda_{HAA}^{} A(m_A^{})
+ \lambda_{Hhh}^{} A(m_{h}^{})
+ 3\lambda_{HHH}^{} A(m_{H}^{})
\right\}.
\end{eqnarray}

\begin{eqnarray}
 \Pi_{hh}^{1PI}(p^2) &=&  
- \frac{N_c}{16\pi^2}
\left[ \frac{m_t^2}{v^2} \frac{\cos^2\al}{\sin^2\be} 
\left\{ 4 A(m_t) + (-2p^2+8 m_t^2) B_0(p^2;m_t,m_t) 
\right\} 
\right. \nonumber\\ && \left.
+
\frac{m_b^2}{v^2} \frac{\sin^2\al}{\cos^2\be} 
\left\{ 4 A(m_b) + (-2p^2+8 m_b^2) B_0(p^2;m_b,m_b) 
\right\}
\right]
 \nonumber\\&&
+ \frac{1}{16\pi^2} \left\{
+(\lambda_{hH^+H^-}^{})^2  B_0(p^2; m_{H^\pm}^{},m_{H^\pm}^{})
+2(\lambda_{hw^+H^-}^{})^2  B_0(p^2; m_{w^\pm},m_{H^\pm}^{})
\right.
\nonumber\\&&
\left.
+(\lambda_{hw^+w^-}^{})^2  B_0(p^2; m_{w^\pm},m_{w^\pm}^{})
- 2 \lambda_{hhH^+H^-} A(m_{H^\pm}^{})
\right.\nonumber\\&&\left.
+(\lambda_{hAA}^{})^2  B_0(p^2; m_{A}^{},m_{A}^{})
+2(\lambda_{hzA}^{})^2  B_0(p^2; m_{z}^{},m_{A}^{})
\right.
\nonumber\\&&
\left.
+(\lambda_{hzz}^{})^2  B_0(p^2; m_{z}^{},m_{z}^{})
- 2 \lambda_{hhAA} A(m_{A}^{})
\right.
\nonumber\\
&&
\left.
+(\lambda_{hHH}^{})^2  B_0(p^2; m_{H}^{},m_{H}^{})
+4(\lambda_{hhH}^{})^2  B_0(p^2; m_{h}^{},m_{H}^{})
\right.
\nonumber\\&&
\left.
+18 (\lambda_{hhh}^{})^2  B_0(p^2; m_{h}^{},m_{h}^{})
- 12 \lambda_{hhhh} A(m_{h}^{})- 2 \lambda_{hhHH} A(m_{H}^{})
 \right\},
\end{eqnarray}
\begin{eqnarray}
 \Pi_{hH}^{1PI}(p^2) &=&  
- \frac{N_c}{16\pi^2} \left[ 
\frac{m_t^2}{v^2} \frac{\sin\al\cos\al}{\sin^2\be} 
\left\{ 4 A(m_t) + (-2p^2+8 m_t^2) B_0(p^2;m_t,m_t) 
\right\}
\right. \nonumber\\ && \left.
- \frac{m_b^2}{v^2} \frac{\sin\al\cos\al}{\sin^2\be} 
\left\{ 4 A(m_b) + (-2p^2+8 m_b^2) B_0(p^2;m_b,m_b) 
\right\}
\right] 
\nonumber\\&&
+ \frac{1}{16\pi^2} \left\{
+\lambda_{hH^+H^-}^{}\lambda_{HH^+H^-}^{}  B_0(p^2; m_{H^\pm}^{},m_{H^\pm}^{})
+2 \lambda_{hw^+H^-}^{}\lambda_{Hw^+H^-}^{}  B_0(p^2; m_{w^\pm},m_{H^\pm}^{})
\right.   
\nonumber\\&&
\left.
+\lambda_{hw^+w^-}^{}\lambda_{Hw^+w^-}^{}  B_0(p^2; m_{w^\pm},m_{w^\pm}^{})
-  \lambda_{hHH^+H^-} A(m_{H^\pm}^{})
\right.\nonumber\\&&\left.
+2 \lambda_{hAA}^{}\lambda_{HAA}^{}  B_0(p^2; m_{A}^{},m_{A}^{})
+\lambda_{hzA}^{}\lambda_{HzA}^{}  B_0(p^2; m_{z}^{},m_{A}^{})
\right.
\nonumber\\&&
\left.
+2 \lambda_{hzz}^{}\lambda_{Hzz}^{}  B_0(p^2; m_{A}^{},m_{A}^{})
- \lambda_{hHAA} A(m_{A}^{})
\right.
\nonumber\\
&&
\left.
+6 \lambda_{hHH}^{}\lambda_{HHH}^{}  B_0(p^2; m_{H}^{},m_{H}^{})
+4 \lambda_{hhH}^{}\lambda_{hHH}^{}  B_0(p^2; m_{h}^{},m_{H}^{})
\right.
\nonumber\\&&
\left.
+6 \lambda_{hhh}^{}\lambda_{hhH}^{}  B_0(p^2; m_{h}^{},m_{h}^{})
- 3 \lambda_{hhhH} A(m_{h}^{})- 3 \lambda_{hHHH} A(m_{H}^{})
 \right\}.
\end{eqnarray}

\begin{eqnarray}
 \Pi_{zA}^{1PI}(p^2) &=&  
- \frac{N_c}{16\pi^2}
 \left[ 
\frac{m_t^2}{v^2} \cot\be 
\left\{ 4 A(m_t) - 2 p^2 B_0(p^2;m_t,m_t) \right\}
\right. \nonumber \\ && \left.
+ \frac{m_b^2}{v^2} \tan\be 
\left\{ 4 A(m_b) -2 p^2 B_0(p^2;m_b,m_b) 
\right\}
\right] 
\nonumber\\&&
+ \frac{1}{16\pi^2} \left\{
-  \lambda_{zAH^+H^-} A(m_{H^\pm}^{})
\right.\nonumber\\&&\left.
+2 \lambda_{HzA}^{}\lambda_{HAA}^{}  B_0(p^2; m_{A}^{},m_{H}^{})
+2 \lambda_{hzA}^{}\lambda_{hAA}^{}  B_0(p^2; m_{A}^{},m_{h}^{})
- 3 \lambda_{zAAA}^{} A(m_{A}^{})
\right.
\nonumber\\&&
\left.
+2 \lambda_{Hzz}^{}\lambda_{HzA}^{}  B_0(p^2; m_{z}^{},m_{H}^{})
+2 \lambda_{hzz}^{}\lambda_{hzA}^{}  B_0(p^2; m_{z}^{},m_{h}^{})
\right.
\nonumber\\&&
\left.
- \lambda_{HHzA} A(m_{H}^{})
- \lambda_{hhzA} A(m_{h}^{})
 \right\}.
\end{eqnarray}

\subsubsection{The $Z$-$A$ mixing} 

The expression of the form factor $\Gamma_{ZA}^{}$ of the $Z$-$A$ mixing 
is given as 
\begin{eqnarray}
\Gamma_{ZA}^{} &=& 
  i \frac{1}{16\pi^2}
  \left\{
 - \frac{m_Z^{}}{v} \cos(\al-\be) \lambda_{hAA}^{} 
   (2 B_1 + B_0)(p^2; m_h, m_A^{})  
\right. 
\nonumber \\
&& 
 - \frac{m_Z^{}}{v} \sin(\al-\be) \lambda_{HAA}^{} 
   (2 B_1 + B_0)(p^2; m_H^{}, m_A^{})  
\nonumber \\
&& 
 + \frac{m_Z^{}}{v} \sin(\al-\be) \lambda_{hzA}^{} 
   (2 B_1 + B_0)(p^2; m_h, m_z)  
\nonumber \\
&& 
 - \frac{m_Z^{}}{v} \cos(\al-\be) \lambda_{HzA}^{} 
   (2 B_1 + B_0)(p^2; m_H, m_z)
\nonumber \\
&& \left.
 + \sum_q N_c \frac{4 m_Z^{} m_q}{v} c_{Aq\bar q} 
    I_q B_0(p^2; m_q, m_q)     
\right\} ,
\end{eqnarray}
where $I_q$ is the iso-spin of the quark $q$,  
$c_{At\bar t}^{}=-\frac{m_t}{v} \cot\beta$ and 
$c_{Ab\bar b}^{}=-\frac{m_b}{v} \tan\beta$ in the Model II THDM, 
and the Higgs self-couplings 
$\lambda_{hAA}^{}$
$\lambda_{HAA}^{}$
$\lambda_{hzA}^{}$ and 
$\lambda_{HzA}^{}$
are listed in Appendix~\ref{app:coupling}.

\subsection{The $hZZ$ vertex} 

The explicit expressions for the 1PI diagrams of 
the form factors of the $hZZ$ vertex are given 
in terms of the Passarino-Veltman functions\cite{pv} by
\begin{eqnarray}
&&M_1^{hZZ}(p_1^2,p_2^2,q^2)= \nonumber\\
&&
+\frac{1}{16\pi^2} \frac{32 N_c m_f^2 m_Z^2}{v^3} c_{hf\bar f}^{}
\left[
\left\{
\frac{1}{2}I_f^2 - I_f Q_f s_W^2 +Q_f^2 s_W^4
\right\} 
\left\{ \frac{}{}
2 p_1^2 C_{21}+2 p_2^2 C_{22} +4 p_1p_2 C_{23} 
\right.\right. \nonumber\\&&
\left.\left.
+ 2 (D-2) C_{24}
+(3p_1^2+p_1p_2) C_{11} 
+ (3p_1p_2 + p_2^2) C_{12} + (p_1^2+p_1p_2) C_0 \frac{}{}
\right\} 
\right. \nonumber \\
&&
\left.
+ \left( I_fQ_fs_W^2-Q_f^2s_W^4 \right) 
\left\{ \frac{}{}
p_1^2 C_{21} + p_2^2 C_{22} +2 p_1p_2 C_{23} + D C_{24} 
\right.\right. \nonumber\\
&& \left.\left.
+ (p_1^2+p_1p_2) C_{11} + (p_1p_2+p_2^2) C_{12} +m_f^2 C_0
\frac{}{}\right\}
\right] (p_1^2,p_2^2,q^2; m_f, m_f, m_f) \nonumber\\
&&
+\frac{1}{16\pi^2} \frac{m_Z^2}{v^2} 
\left[ \frac{}{}
2 \cos^22\theta_W \lambda_{hH^+H^-}^{} 
B_0 (q; m_{H^\pm}^{},m_{H^\pm}^{}) 
\right. \nonumber\\&& \left.
+2 \cos^22\theta_W  \lambda_{hw^+w^-}^{} 
B_0 (q; m_{w^\pm}^{},m_{w^\pm}^{}) 
\right. \nonumber\\&& \left.
+2 \lambda_{hAA}^{} 
B_0 (q; m_{A}^{},m_{A}^{}) 
+
2 \lambda_{hzz}^{} 
B_0 (q; m_{z}^{},m_{z}^{}) 
\right. \nonumber\\&& \left.
+6 \lambda_{hhh}^{} 
B_0 (q; m_{h}^{},m_{h}^{}) 
+
2 \lambda_{hHH}^{} 
B_0 (q; m_{H}^{},m_{H}^{}) 
\right. \nonumber\\&& \left.
-8 \cos^22\theta_W \lambda_{hH^+H^-}^{} 
C_{24} (p_1^2,p_2^2,q^2; m_{H^\pm}^{},m_{H^\pm}^{},m_{H^\pm}^{}) 
\right. \nonumber\\&& \left.
-8 \cos^22\theta_W \lambda_{hw^+w^-}^{} 
C_{24} (p_1^2,p_2^2,q^2; m_{w^\pm}^{},m_{w^\pm}^{},m_{w^\pm}^{}) 
\right. \nonumber\\&& \left.
-8 \sin^2(\al-\be) 
\left\{ \frac{}{}
\lambda_{hzz}^{} 
C_{24} (p_1^2,p_2^2,q^2; m_{z}^{},m_{h}^{},m_{z}^{}) 
+\lambda_{hAA}^{} 
C_{24} (p_1^2,p_2^2,q^2; m_{A}^{},m_{H}^{},m_{A}^{}) \right.
\right. \nonumber\\&& \left. 
\left.
+\lambda_{hHH}^{} 
C_{24} (p_1^2,p_2^2,q^2; m_{H}^{},m_{A}^{},m_{H}^{}) 
+3 \lambda_{hhh}^{} 
C_{24} (p_1^2,p_2^2,q^2; m_{h}^{},m_{z}^{},m_{h}^{}) \frac{}{}\right\}
\right. \nonumber\\&& \left. 
-8 \cos^2(\al-\be) 
\left\{ \frac{}{}
\lambda_{hAA}^{} 
C_{24} (p_1^2,p_2^2,q^2; m_{A}^{},m_{h}^{},m_{A}^{}) 
+\lambda_{hzz}^{} 
C_{24} (p_1^2,p_2^2,q^2; m_{z}^{},m_{H}^{},m_{z}^{}) \right.
\right. \nonumber\\&& \left. 
\left.
+\lambda_{hHH}^{} 
C_{24} (p_1^2,p_2^2,q^2; m_{H}^{},m_{z}^{},m_{H}^{}) 
+3 \lambda_{hhh}^{} 
C_{24} (p_1^2,p_2^2,q^2; m_{h}^{},m_{A}^{},m_{h}^{}) \frac{}{}\right\}
\right. \nonumber\\&& \left. 
+4 \cos(\al-\be) \sin(\al-\be) 
\left\{ \frac{}{}
\lambda_{hzA}^{} 
C_{24} (p_1^2,p_2^2,q^2; m_{z}^{},m_{h}^{},m_{A}^{}) 
+\lambda_{hzA}^{} 
C_{24} (p_1^2,p_2^2,q^2; m_{A}^{},m_{h}^{},m_{z}^{}) \right.
\right. \nonumber\\&& \left. 
\left.
-\lambda_{hzA}^{} 
C_{24} (p_1^2,p_2^2,q^2; m_{z}^{},m_{H}^{},m_{A}^{}) 
-\lambda_{hzA}^{} 
C_{24} (p_1^2,p_2^2,q^2; m_{A}^{},m_{H}^{},m_{z}^{}) \right.
\right. \nonumber\\&& \left. 
\left.
+2 \lambda_{hhH}^{} 
C_{24} (p_1^2,p_2^2,q^2; m_{h}^{},m_{z}^{},m_{H}^{}) 
+2 \lambda_{hhH}^{} 
C_{24} (p_1^2,p_2^2,q^2; m_{H}^{},m_{z}^{},m_{h}^{}) \right.
\right. \nonumber\\&& \left. 
\left.
-2 \lambda_{hhH}^{} 
C_{24} (p_1^2,p_2^2,q^2; m_{h}^{},m_{A}^{},m_{H}^{}) 
-2 \lambda_{hhH}^{} 
C_{24} (p_1^2,p_2^2,q^2; m_{H}^{},m_{A}^{},m_{h}^{}) 
\frac{}{}\right\}
\frac{}{}\right], 
\end{eqnarray}
\begin{eqnarray}
&&M_2^{hZZ}(p_1^2,p_2^2,q^2)= 
\nonumber\\&&
-\frac{1}{16\pi^2} \frac{32 N_c m_f^2 m_Z^4}{v^3} c_{hf\bar f}^{}
\left[
\left\{
\frac{1}{2}I_f^2 - I_f Q_f s_W^2 +Q_f^2 s_W^4
\right\} 
\left\{ \frac{}{}
4 C_{23}+3 C_{12} + C_{11} + C_0 \right\} 
\right. \nonumber \\
&&
\left.
+ \left( I_fQ_fs_W^2-Q_f^2s_W^4 \right) 
\left\{ \frac{}{}
 C_{11} -  C_{12} \right\}
\right] (p_1^2,p_2^2,q^2; m_f, m_f, m_f)\nonumber\\
&&
+\frac{1}{16\pi^2} \frac{m_Z^4}{v^2} 
\left[ \frac{}{}
-8 \cos^22\theta_W \lambda_{hH^+H^-}^{} 
C_{1223} (p_1^2,p_2^2,q^2; m_{H^\pm}^{},m_{H^\pm}^{},m_{H^\pm}^{}) 
\right. \nonumber\\&& \left.
-8 \cos^22\theta_W \lambda_{hw^+w^-}^{} 
C_{1223} (p_1^2,p_2^2,q^2; m_{w^\pm}^{},m_{w^\pm}^{},m_{w^\pm}^{}) 
\right. \nonumber\\&& \left.
-8 \sin^2(\al-\be) 
\left\{ \frac{}{}
\lambda_{hzz}^{} 
C_{1223} (p_1^2,p_2^2,q^2; m_{z}^{},m_{h}^{},m_{z}^{}) 
+\lambda_{hAA}^{} 
C_{1223} (p_1^2,p_2^2,q^2; m_{A}^{},m_{H}^{},m_{A}^{}) \right.
\right. \nonumber\\&& \left. 
\left.
+\lambda_{hHH}^{} 
C_{1223} (p_1^2,p_2^2,q^2; m_{H}^{},m_{A}^{},m_{H}^{}) 
+3 \lambda_{hhh}^{} 
C_{1223} (p_1^2,p_2^2,q^2; m_{h}^{},m_{z}^{},m_{h}^{}) \frac{}{}\right\}
\right. \nonumber\\&& \left. 
-8 \cos^2(\al-\be) 
\left\{ \frac{}{}
\lambda_{hAA}^{} 
C_{1223} (p_1^2,p_2^2,q^2; m_{A}^{},m_{h}^{},m_{A}^{}) 
+\lambda_{hzz}^{} 
C_{1223} (p_1^2,p_2^2,q^2; m_{z}^{},m_{H}^{},m_{z}^{}) \right.
\right. \nonumber\\&& \left. 
\left.
+\lambda_{hHH}^{} 
C_{1223} (p_1^2,p_2^2,q^2; m_{H}^{},m_{z}^{},m_{H}^{}) 
+3 \lambda_{hhh}^{} 
C_{1223} (p_1^2,p_2^2,q^2; m_{h}^{},m_{A}^{},m_{h}^{}) \frac{}{}\right\}
\right. \nonumber\\&& \left. 
+4 \cos(\al-\be) \sin(\al-\be) 
\left\{ \frac{}{}
\lambda_{hzA}^{} 
C_{1223} (p_1^2,p_2^2,q^2; m_{z}^{},m_{h}^{},m_{A}^{}) 
+\lambda_{hzA}^{} 
C_{1223} (p_1^2,p_2^2,q^2; m_{A}^{},m_{h}^{},m_{z}^{}) \right.
\right. \nonumber\\&& \left. 
\left.
-\lambda_{hzA}^{} 
C_{1223} (p_1^2,p_2^2,q^2; m_{z}^{},m_{H}^{},m_{A}^{}) 
-\lambda_{hzA}^{} 
C_{1223} (p_1^2,p_2^2,q^2; m_{A}^{},m_{H}^{},m_{z}^{}) \right.
\right. \nonumber\\&& \left. 
\left.
+2 \lambda_{hhH}^{} 
C_{1223} (p_1^2,p_2^2,q^2; m_{h}^{},m_{z}^{},m_{H}^{}) 
+2 \lambda_{hhH}^{} 
C_{1223} (p_1^2,p_2^2,q^2; m_{H}^{},m_{z}^{},m_{h}^{}) \right.
\right. \nonumber\\&& \left. 
\left.
-2 \lambda_{hhH}^{} 
C_{1223} (p_1^2,p_2^2,q^2; m_{h}^{},m_{A}^{},m_{H}^{}) 
-2 \lambda_{hhH}^{} 
C_{1223} (p_1^2,p_2^2,q^2; m_{H}^{},m_{A}^{},m_{h}^{}) 
\frac{}{}\right\}
\frac{}{}\right], 
\end{eqnarray}
\begin{eqnarray}
&&M_3^{hZZ}(p_1^2,p_2^2,q^2)= 
\nonumber\\&&
-\frac{1}{16\pi^2} \frac{32 N_c m_f^2 m_Z^4}{v^3} c_{hf\bar f}^{}
\left( \frac{1}{2}I_f^2 - I_f Q_fs_W^2\right) 
\left\{ \frac{}{}
 C_{12} -  C_{11} - C_0 \right\}
(p_1^2,p_2^2,q^2; m_f, m_f, m_f), \nonumber \\
\end{eqnarray}
where $c_{ht\bar t}=\cos\al/\sin\be$ and
      $c_{hb\bar b}=-\sin\al/\cos\be$ in the Model II THDM, and 
the each coupling constant of Higgs bosons is 
listed in Appendix~\ref{app:coupling}.

\subsection{The 1PI $hhh$ form factor}

The explicit expression for the 1PI diagrams 
of the $hhh$ form factor is given 
in terms of the Passarino-Veltman functions\cite{pv} by
\begin{eqnarray}
&&\Gamma_{hhh}^{1PI}(p_1^2,p_2^2,q^2)=\nonumber\\
&&-\!\!\sum_{f=t,b} \left[
\frac{1}{16\pi^2} \frac{8 N_c m_f^4}{v^3} 
c_{hf\bar f}^3
\left\{ \frac{}{}
3 \left(
p_1^2 C_{21} + p_2^2 C_{22} 
+2 p_1p_2 C_{23} + D C_{24}
\right) 
+\left( 4 p_1^2 + 2 p_1p_2 \right) C_{11}
\right. \right. 
\nonumber\\
&& \left.\left.
+\left( 2 p_2^2 + 4 p_1p_2 \right) C_{12} 
+ \left(m_f^2 + p_1^2+p_1p_2\right) C_0
\right\} \right]
(p_1^2,p_2^2,q^2; m_f, m_{f}, m_f)
\nonumber\\&& 
+ \frac{1}{16\pi^2} 
\left[ 
2 \lambda_{hH^+H^-}^{}\lambda_{hhH^+H^-}^{} 
\left\{\frac{}{}
B_0 (q^2; m_{H^\pm}^{},m_{H^\pm}^{}) 
+B_0 (p_1^2; m_{H^\pm}^{},m_{H^\pm}^{})
\right. \right. \nonumber \\&& \left. \left. 
+B_0 (p_2^2; m_{H^\pm}^{},m_{H^\pm}^{}) 
\frac{}{}\right\} 
\right.\nonumber \\&& \left.
+4 \lambda_{hw^+H^-}^{}\lambda_{hhw^+H^-}^{} 
\left\{\frac{}{}
B_0 (q^2; m_{w^\pm}^{},m_{H^\pm}^{}) 
+B_0 (p_1^2; m_{w^\pm}^{},m_{H^\pm}^{})
+B_0 (p_2^2; m_{w^\pm}^{},m_{H^\pm}^{}) 
\frac{}{}\right\} 
\right.\nonumber \\&& \left.
+2 \lambda_{hw^+w^-}^{}\lambda_{hhw^+w^-}^{} 
\left\{\frac{}{}
B_0 (q^2; m_{w^\pm}^{},m_{w^\pm}^{}) 
+B_0 (p_1^2; m_{w^\pm}^{},m_{w^\pm}^{})
+B_0 (p_2^2; m_{w^\pm}^{},m_{w^\pm}^{}) 
\frac{}{}\right\} 
\right.\nonumber \\&& \left.
-2 \lambda_{hH^+H^-}^{3} 
C_0 (p_1^2,p_2^2,q^2; m_{H^\pm}^{},m_{H^\pm}^{},m_{H^\pm}^{})
\right.\nonumber \\&& \left.
-2 \lambda_{hH^+H^-}^{} \lambda_{hw^+H^-}^{2}
\left\{\frac{}{}
C_0 (p_1^2,p_2^2,q^2; m_{H^\pm}^{},m_{H^\pm}^{},m_{w^\pm}^{})
\right.
\right.\nonumber \\&& \left.
\left.
+C_0 (p_1^2,p_2^2,q^2; m_{H^\pm}^{},m_{w^\pm}^{},m_{H^\pm}^{})
+C_0 (p_1^2,p_2^2,q^2; m_{w^\pm}^{},m_{H^\pm}^{},m_{H^\pm}^{})
\frac{}{}\right\} 
\right.\nonumber \\&& \left.
-2 \lambda_{hw^+w^-}^{} \lambda_{hw^+H^-}^{2}
\left\{\frac{}{}
C_0 (p_1^2,p_2^2,q^2; m_{H^\pm}^{},m_{w^\pm}^{},m_{w^\pm}^{})
\right.
\right.\nonumber \\&& \left.
\left.
+C_0 (p_1^2,p_2^2,q^2; m_{w^\pm}^{},m_{H^\pm}^{},m_{w^\pm}^{})
+C_0 (p_1^2,p_2^2,q^2; m_{w^\pm}^{},m_{w^\pm}^{},m_{H^\pm}^{})
\frac{}{}\right\} 
\right.\nonumber \\&& \left.
-2 \lambda_{hw^+w^-}^{3} 
C_0 (p_1^2,p_2^2,q^2; m_{w^\pm}^{},m_{w^\pm}^{},m_{w^\pm}^{})
\right.\nonumber \\&& \left.
+4 \lambda_{hAA}^{}\lambda_{hhAA}^{} 
\left\{\frac{}{}
 B_0 (q^2; m_{A}^{},m_{A}^{}) 
+B_0 (p_1^2; m_{A}^{},m_{A}^{})
+B_0 (p_2^2; m_{A}^{},m_{A}^{}) 
\frac{}{}\right\} 
\right.\nonumber \\&& \left.
+2 \lambda_{hzA}^{}\lambda_{hhzA}^{} 
\left\{\frac{}{}
 B_0 (q^2; m_{z}^{},m_{A}^{}) 
+B_0 (p_1^2;m_{z}^{},m_{A}^{})
+B_0 (p_2^2;m_{z}^{},m_{A}^{}) 
\frac{}{}\right\} 
\right.\nonumber \\&& \left.
+4 \lambda_{hzz}^{}\lambda_{hhzz}^{} 
\left\{\frac{}{}
 B_0 (q^2; m_{z}^{},m_{z}^{}) 
+B_0 (p_1^2; m_{z}^{},m_{z}^{}) 
+B_0 (p_2^2; m_{z}^{},m_{z}^{}) 
\frac{}{}\right\} 
\right.\nonumber \\&& \left.
-4 \lambda_{hAA}^{3} 
C_0 (p_1^2,p_2^2,q^2; m_{A}^{},m_{A}^{},m_{A}^{})
\right.\nonumber \\&& \left.
-2 \lambda_{hAA}^{} \lambda_{hzA}^{2}
\left\{\frac{}{}
C_0 (p_1^2,p_2^2,q^2; m_{A}^{},m_{A}^{},m_{z}^{})
\right.
\right.\nonumber \\&& \left.
\left.
+C_0 (p_1^2,p_2^2,q^2; m_{A}^{},m_{z}^{},m_{A}^{})
+C_0 (p_1^2,p_2^2,q^2; m_{z}^{},m_{A}^{},m_{A}^{})
\frac{}{}\right\} 
\right.\nonumber \\&& \left.
-2 \lambda_{hzz}^{} \lambda_{hzA}^{2}
\left\{\frac{}{}
C_0 (p_1^2,p_2^2,q^2; m_{A}^{},m_{z}^{},m_{z}^{})
\right.
\right.\nonumber \\&& \left.
\left.
+C_0 (p_1^2,p_2^2,q^2; m_{z}^{},m_{A}^{},m_{z}^{})
+C_0 (p_1^2,p_2^2,q^2; m_{z}^{},m_{z}^{},m_{A}^{})
\frac{}{}\right\} 
\right.\nonumber \\&& \left.
-4 \lambda_{hzz}^{3} 
C_0 (p_1^2,p_2^2,q^2; m_{z}^{},m_{z}^{},m_{z}^{})
\right.\nonumber \\&& \left.
+4 \lambda_{hHH}^{}\lambda_{hhHH}^{} 
\left\{\frac{}{}
 B_0 (q^2; m_{H}^{},m_{H}^{}) 
+B_0 (p_1^2; m_{H}^{},m_{H}^{})
+B_0 (p_2^2; m_{H}^{},m_{H}^{}) 
\frac{}{}\right\} 
\right.\nonumber \\&& \left.
+12 \lambda_{hhH}^{}\lambda_{hhhH}^{} 
\left\{\frac{}{}
 B_0 (q^2;  m_{h}^{},m_{H}^{}) 
+B_0 (p_1^2;m_{h}^{},m_{H}^{})
+B_0 (p_2^2;m_{h}^{},m_{H}^{}) 
\frac{}{}\right\} 
\right.\nonumber \\&& \left.
+72 \lambda_{hhh}^{}\lambda_{hhhh}^{} 
\left\{\frac{}{}
 B_0 (q^2; m_{h}^{},m_{h}^{}) 
+B_0 (p_1^2; m_{h}^{},m_{h}^{}) 
+B_0 (p_2^2; m_{h}^{},m_{h}^{}) 
\frac{}{}\right\} 
\right.\nonumber \\&& \left.
-108 \lambda_{hhh}^{3} 
C_0 (p_1^2,p_2^2,q^2; m_{h}^{},m_{h}^{},m_{h}^{})
\right.\nonumber \\&& \left.
-24 \lambda_{hhh}^{} \lambda_{hhH}^{2}
\left\{\frac{}{}
C_0 (p_1^2,p_2^2,q^2; m_{H}^{},m_{h}^{},m_{h}^{})
\right.
\right.\nonumber \\&& \left.
\left.
+C_0 (p_1^2,p_2^2,q^2; m_{h}^{},m_{H}^{},m_{h}^{})
+C_0 (p_1^2,p_2^2,q^2; m_{h}^{},m_{h}^{},m_{H}^{})
\frac{}{}\right\} 
\right.\nonumber \\&& \left.
-8 \lambda_{hHH}^{} \lambda_{hhH}^{2}
\left\{\frac{}{}
C_0 (p_1^2,p_2^2,q^2; m_{H}^{},m_{h}^{},m_{H}^{})
\right.
\right.\nonumber \\&& \left.
\left.
+C_0 (p_1^2,p_2^2,q^2; m_{H}^{},m_{H}^{},m_{h}^{})
+C_0 (p_1^2,p_2^2,q^2; m_{h}^{},m_{H}^{},m_{H}^{})
\frac{}{}\right\} 
\right.\nonumber \\&& \left.
-4 \lambda_{hHH}^{3} 
C_0 (p_1^2,p_2^2,q^2; m_{H}^{},m_{H}^{},m_{H}^{})
\right], 
\end{eqnarray}
where $c_{ht\bar t}=\cos\al/\sin\be$ and
      $c_{hb\bar b}=-\sin\al/\cos\be$ in the Model II THDM, and 
the each coupling constant of Higgs bosons is 
listed in Appendix~\ref{app:coupling}.

\section{Wave function renormalization}\label{app:reno-mixing}

Let us consider the wave functions of $h$ and $H$. 
The bare scalars satisfy 
\begin{eqnarray}
\left[  \begin{array}{c} {h_1}_B^{} \\ {h_2}_B^{} \\
        \end{array}\right]
=
\left[  \begin{array}{cc} \cos\al_B^{} & -\sin\al_B^{} \\ 
                          \sin\al_B^{} &  \cos\al_B^{} \\
        \end{array}\right]
\left[  \begin{array}{c} H_B^{} \\ h_B^{} \\
        \end{array}\right]
=
R(\al_B^{})
\left[  \begin{array}{c} H_B^{} \\ h_B^{} \\
        \end{array}\right].
\end{eqnarray}
We can write 
\begin{eqnarray}
\left[  \begin{array}{c} H_B^{} \\ h_B^{} \\
        \end{array}\right]
&=&
R(-\al_B^{})
\left[  \begin{array}{c} {h_1}_B^{} \\ {h_2}_B^{} \\
        \end{array}\right] 
=
R(-\delta \al) R (- \al)
\left[  \begin{array}{c} {h_1}_B^{} \\ {h_2}_B^{} \\
        \end{array}\right] \nonumber\\
&\to&
R(-\delta \al) R (- \al)
\tilde{Z}
\left[  \begin{array}{c} {h_1} \\ {h_2} \\
        \end{array}\right] 
=
R(-\delta \al) 
Z
\left[  \begin{array}{c} H \\ h \\
        \end{array}\right], 
\end{eqnarray}
where $\tilde{Z}$ is a arbitrary real symmetric matrix, so that  
\begin{eqnarray}
Z= R(-\al) \tilde{Z} R(\al) 
\equiv 
\left[  \begin{array}{cc} Z_{HH}^{1/2} & Z_{Hh}^{1/2} \\ 
                          Z_{hH}^{1/2} & Z_{hh}^{1/2} \\
        \end{array}\right] 
\end{eqnarray}
is also arbitrary symmetric ($Z_{hH}=Z_{Hh}$). 
We may expand these elements by 
\begin{eqnarray}
 Z_{HH}^{} &=& 1 + \delta Z_{H}^{} + \cdot \cdot \cdot , \\
 Z_{hh}^{} &=& 1 + \delta Z_{h}^{} + \cdot \cdot \cdot , \\
 Z_{Hh}^{} &=& 0 + \delta C_h +     \cdot \cdot \cdot . 
\end{eqnarray}
In this way, we obtain Eq.~(\ref{mixwave1}).

\section{Perturbative Unitarity}

The condition of perturbative unitarity has originally been 
discussed for the elastic scattering of the 
longitudinally polarized gauge bosons and the Higgs boson 
by Lee, Quigg and Thacker\cite{LQT}.   
The channels 
$W^+_LW^-_L$, $Z_L^{}Z_L^{}$, $Z_L^{} h$, $hh$
are considered as the initial and final states, and 
the condition in Eq.~(\ref{swaveu}) with $\xi=1$ is imposed to each 
eigenvalue of the $4 \times 4$ S-matrix. 

The extension to the THDM has been studied by several 
authors\cite{unitarity1,unitarity2}. 
The 14 channels, 
\begin{eqnarray} 
W^+_LW^-_L, W^+_LH^-, H^+W^-_L, H^+H^-, 
Z_L^{}Z_L^{}, Z_L^{} A, AA,
Z_L^{} h, Z_L^{} H, A h, A H, 
hh, hH,  HH,  
\end{eqnarray} 
have been taken into account, 
and the equivalence theorem\cite{et,et2} has been 
employed to evaluate the tree level S-wave amplitudes 
for each channel in Ref.~\cite{unitarity1}. 
The 14 eigenvalues of the S-matrix are calculated
and their expressions are given in terms of the 
Higgs coupling constants by 
\begin{eqnarray} 
a_{\pm}^{} &=& \frac{1}{16\pi} 
  \left\{
  \frac{3}{2} (\lambda_1 + \lambda_2) \pm 
   \sqrt{\frac{9}{4} (\lambda_1-\lambda_2)^2 + (2\lambda_3+\lambda_4)^2}
  \right\},  \\
b_{\pm}^{} &=& \frac{1}{16\pi} 
  \left\{
  \frac{1}{2} (\lambda_1 + \lambda_2) \pm 
   \sqrt{\frac{1}{4} (\lambda_1-\lambda_2)^2 + \lambda_4^2}
  \right\},  \\
c_{\pm}^{} &=& d_{\pm}^{}= \frac{1}{16\pi} 
  \left\{
  \frac{1}{2} (\lambda_1 + \lambda_2) \pm 
   \sqrt{\frac{1}{4} (\lambda_1-\lambda_2)^2 + \lambda_5^2}
  \right\},  \\
e_1 &=&  \frac{1}{16\pi}  \left(
    \lambda_3 + 2 \lambda_4 - 3 \lambda_5 \right), \\
e_2 &=&  \frac{1}{16\pi}  \left(
    \lambda_3 - \lambda_5 \right), \\
f_+ &=&  \frac{1}{16\pi}  \left(
    \lambda_3 + 2 \lambda_4 + 3 \lambda_5 \right), \\
f_- &=&  \frac{1}{16\pi}  \left(
    \lambda_3 + \lambda_5 \right), \\
f_1 &=& f_2 =  \frac{1}{16\pi}  \left(
    \lambda_3 + \lambda_4 \right).  
\end{eqnarray} 
The perturbative unitarity condition is then expressed by 
\begin{eqnarray}
  |a_{\pm}|,  |b_{\pm}|,  |c_{\pm}|,  |d_{\pm}|,  
   |e_{1,2}^{}|,  |f_{\pm}|,  |f_{1,2}^{}|  < \xi. \label{pvu_THDM}
\end{eqnarray}
The condition in (\ref{pvu_THDM}) with the tree level mass formulas 
in Eqs.~(\ref{mass_THDM1}) to (\ref{mass_THDM5}) 
constrains the parameter space of the Higgs sector. 
The parameter $\xi$ is taken to be $1/2$ in our numerical 
evaluation\cite{HHG}.

\section{Higgs couplings in the THDM}\label{app:coupling}

Here, we list the Higgs boson self-coupling constants  in the THDM,
which are expressed in terms of our input parameters.

\subsection{Trilinear Higgs couplings}

\begin{eqnarray}
\nonumber\lambda_{hhh}&=&
-\frac{1}{4v\sin2\beta}\Big[\Big\{\cos(3\alpha-\beta)+3\cos(\alpha+\beta)\Big\}m_h^2\\
&&\hspace{5cm}-4\cos^2(\alpha-\beta)\cos(\alpha+\beta)M^2\Big],\\
\lambda_{hHH}&=&-\frac{\sin(\alpha-\beta)}{2v\sin2\beta}\bigg[\sin2\alpha(m_h^2+2m_H^2)-
(3\sin2\alpha+\sin2\beta)M^2\bigg],\\
\lambda_{Hhh}&=&-\frac{\cos(\alpha-\beta)}{2v\sin2\beta}\bigg[\sin2\alpha(2m_h^2+m_H^2)-(3\sin2\alpha-\sin2\beta)M^2\bigg],\\
\nonumber\lambda_{HHH}&=&\frac{1}{4v\sin2\beta}\Big[\Big\{\sin(3\alpha-\beta)-3\sin(\alpha+\beta)\Big\}m_H^2\\
&&\hspace{5cm}+4\sin^2(\alpha-\beta)\sin(\alpha+\beta)M^2\Big],\\
\nonumber\lambda_{hAA}&=&-\frac{1}{4v\sin2\beta}\Big[\Big\{\cos(\alpha-3\beta)+3\cos(\alpha+\beta)\Big\}m_h^2\\
&&\hspace{2cm}-4\sin2\beta\sin(\alpha-\beta)m_A^2-4\cos(\alpha+\beta)M^2\Big],\\
\nonumber\lambda_{HAA}&=&-\frac{1}{4v\sin2\beta}\Big[\Big\{\sin(\alpha-3\beta)+3\sin(\alpha+\beta)\Big\}m_H^2\\
&&\hspace{2cm}+4\sin2\beta\cos(\alpha-\beta)m_A^2-4\sin(\alpha+\beta)M^2\Big],\\
\lambda_{hzz}&=&\frac{m_h^2}{2v}\sin(\alpha-\beta),\\
\lambda_{Hzz}&=&-\frac{m_H^2}{2v}\cos(\alpha-\beta),\\
\lambda_{hzA}&=&\frac{m_A^2-m_h^2}{v}\cos(\alpha-\beta),\\
\lambda_{HzA}&=&\frac{m_A^2-m_H^2}{v}\sin(\alpha-\beta),\\
\nonumber\lambda_{hH^+H^-}&=&-\frac{1}{2v\sin2\beta}\Big[\Big\{\cos(\alpha-3\beta)+3\cos(\alpha+\beta)\Big\}m_h^2\\
&&\hspace{2cm}-4\sin2\beta\sin(\alpha-\beta)m_{H^\pm}^2-4\cos(\alpha+\beta)M^2\Big],\\
\nonumber\lambda_{HH^+H^-}&=&-\frac{1}{2v\sin2\beta}\Big[\Big\{\sin(\alpha-3\beta)+3\sin(\alpha+\beta)\Big\}m_H^2\\
&&\hspace{2cm}+4\sin2\beta\cos(\alpha-\beta)m_{H^\pm}^2-4\sin(\alpha+\beta)M^2\Big],\\
\lambda_{hw^+w^-}&=&\frac{m_h^2}{v}\sin(\alpha-\beta),\\
\lambda_{Hw^+w^-}&=&-\frac{m_H^2}{v}\cos(\alpha-\beta),\\
\lambda_{hw^+ H^-}&=&\frac{m_{H^\pm}^2-m_h^2}{v}\cos(\alpha-\beta),\\
\lambda_{Hw^+ H^-}&=&\frac{m_{H^\pm}^2-m_H^2}{v}\sin(\alpha-\beta),
\end{eqnarray}

\subsection{Quartic couplings}

\begin{eqnarray}
\nonumber\lambda_{hhhh}&=&-\frac{1}{32v^2\sin^22\beta}\Big[\Big\{\cos(3\alpha-\beta)+3\cos(\alpha+\beta)\Big\}^2m_h^2\\
&&\hspace{2.5cm}+4\sin^22\alpha\cos^2(\alpha-\beta)m_H^2-4(\cos2\alpha+\cos2\beta)^2M^2\Big],\\
\nonumber\lambda_{hhhH}&=&-\frac{\sin2\alpha\cos(\alpha-\beta)}{4v^2\sin^22\beta}\Big[\Big\{\cos(3\alpha-\beta)+3\cos(\alpha+\beta)\Big\}m_h^2\\
&&\hspace{3.5cm}+2\sin2\alpha\sin(\alpha-\beta)m_H^2-4\cos(\alpha+\beta)M^2\Big],\\
\nonumber\lambda_{hhHH}&=&-\frac{1}{16v^2\sin^22\beta}\Big[\sin2\alpha\Big\{6\sin2\alpha+3\sin(4\alpha-2\beta)-\sin2\beta\Big\}m_h^2\\
\nonumber&&\hspace{2.5cm}+\sin2\alpha\Big\{6\sin2\alpha-3\sin(4\alpha-2\beta)+\sin2\beta\Big\}m_H^2\\
&&\hspace{6cm}-2(2-3\cos4\alpha+\cos4\beta)M^2\Big],\\
\nonumber\lambda_{zAAA}&=&-\frac{1}{8v^2\sin2\beta}\Big[\Big\{3\cos2\alpha+\cos(2\alpha-4\beta)+4\cos2\beta\Big\}m_h^2\\
&&\hspace{2cm}-\Big\{3\cos2\alpha+\cos(2\alpha-4\beta)-4\cos2\beta\Big\}m_H^2-8\cos2\beta M^2\Big],\\
\nonumber\lambda_{hhAA}&=&-\frac{1}{32v^{2}\sin^{2}2\beta}\Big[\Big\{9+3\cos4\alpha+6\cos(2\alpha-2\beta)\\
\nonumber&&\hspace{2.5cm}+\cos(4\alpha-4\beta)+3\cos4\beta+10\cos(2\alpha+2\beta)\Big\}m_h^2\\
\nonumber&&\hspace{2.5cm}+2\sin2\beta\Big\{3\sin2\alpha+\sin(2\alpha-4\beta)+2\sin2\beta\Big\}m_H^2\\
\nonumber&&\hspace{2.5cm}+16\sin^{2}2\beta\sin^{2}(\alpha-\beta)m_A^2\\
\nonumber&&\hspace{2.5cm}+2\Big\{6+\cos(2\alpha-6\beta)+2\cos(2\alpha-2\beta)\\
&&\hspace{5.5cm}+2\cos4\beta+5\cos(2\alpha+2\beta)\Big\}M^2\Big],\\
\nonumber\lambda_{hhzz}&=&\frac{1}{16v^2\sin2\beta}\Big[\Big\{2\sin2\alpha+\sin(4\alpha-2\beta)-3\sin2\beta)\Big\}m_h^2\\
&&\qquad-4\sin2\alpha\cos^2(\alpha-\beta)m_H^2-8\sin2\beta\cos^2(\alpha-\beta)(m_A^2-M^2)\Big],\\
\nonumber\lambda_{hhzA}&=&-\frac{\cos(\alpha-\beta)}{4v^2\sin2\beta}\Big[\Big\{\cos(3\alpha-\beta)+3\cos(\alpha+\beta)\Big\}m_h^2+2\sin2\alpha\sin(\alpha-\beta)m_H^2\\
&&\hspace{3cm}+4\sin2\beta\sin(\alpha-\beta)m_A^2-4\cos2\beta\cos(\alpha-\beta)M^2\Big],\\
\nonumber\lambda_{HHzA}&=&-\frac{\sin(\alpha-\beta)}{4v^2\sin2\beta}\Big[2\sin2\alpha\cos(\alpha-\beta)m_h^2-\Big\{\sin(3\alpha-\beta)-3\sin(\alpha+\beta)\Big\}m_H^2\\
&&\hspace{3cm}-4\sin2\beta\cos(\alpha-\beta)m_A^2-4\cos2\beta\sin(\alpha-\beta)M^2\Big],\\
\nonumber\lambda_{hhH^+H^-}&=&-\frac{1}{16v^2\sin^22\beta}\Big[\Big\{9+3\cos4\alpha+6\cos(2\alpha-2\beta)\\
\nonumber&&\hspace{3.7cm}+\cos(4\alpha-4\beta)+3\cos4\beta+10\cos(2\alpha+2\beta)\Big\}m_h^2\\
\nonumber&&\hspace{2.3cm}+\Big\{3-3\cos4\alpha+2\cos(2\alpha-2\beta)\\
\nonumber&&\hspace{3.7cm}-\cos(4\alpha-4\beta)+\cos4\beta-2\cos(2\alpha+2\beta)\Big\}m_H^2\\
\nonumber&&\hspace{2.5cm}+16\sin^2(\alpha-\beta)\sin^22\beta m_{H^\pm}^2\\
\nonumber&&\hspace{2.5cm}-2\Big\{6+\cos(2\alpha-6\beta)+2\cos(2\alpha-2\beta)\\
&&\hspace{5.5cm}+2\cos4\beta+5\cos(2\alpha+2\beta)\Big\}M^2\Big],\\
\nonumber\lambda_{hhw^+w^-}&=&\frac{1}{8v^2\sin2\beta}\Big[\Big\{2\sin2\alpha+\sin(4\alpha-2\beta)-3\sin2\beta\Big\}m_h^2-4\sin2\alpha\cos^2(\alpha-\beta)m_H^2\\
&&\hspace{5cm}-8\sin2\beta\cos^2(\alpha-\beta)(m_{H^\pm}^2-M^2)\Big],\\
\nonumber\lambda_{hhw^+ H^-}&=&-\frac{\cos(\alpha-\beta)}{4v^2\sin2\beta}\Big[\Big\{\cos(3\alpha-\beta)+3\cos(\alpha+\beta)\Big\}m_h^2+2\sin2\alpha\sin(\alpha-\beta)m_H^2\\
&&\hspace{2.5cm}+4\sin2\beta\sin(\alpha-\beta)m_{H^\pm}^2-4\cos2\beta\cos(\alpha-\beta)M^2\Big],\\
 \nonumber\lambda_{hHH^+H^-}&=&-\frac{1}{4v^2\sin^22\beta}\Big[\sin2\alpha\Big\{3\cos2\alpha+\cos(2\alpha-4\beta)+4\cos2\beta\Big\}m_h^2\\
\nonumber&&\hspace{2.5cm}+\sin2\alpha\Big\{-3\cos2\alpha-\cos(2\alpha-4\beta)+4\cos2\beta\Big\}m_H^2\\
\nonumber&&\hspace{2.5cm}-4\sin^22\beta\sin(2\alpha-2\beta)m_{H^\pm}^2\\
&&\hspace{2.5cm}-4\Big\{2\sin2\alpha\cos^32\beta+\sin^22\beta\sin(2\alpha+2\beta)\Big\}M^2\Big],\\
\nonumber\lambda_{zAH^+H^-}&=&-\frac{1}{4v^2\sin2\beta}\Big[\Big\{3\cos2\alpha+\cos(2\alpha-4\beta)+4\cos2\beta\Big\}m_h^2\\
&&\hspace{2cm}-\Big\{3\cos2\alpha+\cos(2\alpha-4\beta)-4\cos2\beta\Big\}m_H^2-8\cos2\beta M^2\Big].
\end{eqnarray}


\end{document}